\documentclass[12pt]{article}
\usepackage{amssymb,amsmath,epsfig,xcolor}
\allowdisplaybreaks

\begin{document}

\title{\bf Role of Non-conserved Gravity Theory and Electric Charge in Constructing Complexity-free
Stellar Models: A Novel Approach under Non-minimal Coupling}
\author{Tayyab Naseer$^{1,2}$ \thanks{tayyab.naseer@math.uol.edu.pk; tayyabnaseer48@yahoo.com}\\
$^1$Department of Mathematics and Statistics, The University of Lahore,\\
1-KM Defence Road Lahore-54000, Pakistan.\\
$^2$Research Center of Astrophysics and Cosmology, Khazar University, \\
Baku, AZ1096, 41 Mehseti Street, Azerbaijan.}

\date{}
\maketitle

\begin{abstract}
This study explores the application of complexity factor within the
context of Rastall gravity, exploring its implications on a static
spacetime admitting spherical symmetry associated with anisotropic
fluids under an electromagnetic field. The field equations are
derived for a static charged sphere that provides a foundational
framework for analyzing gravitational effects in this non-conserved
theory. The mass function is formulated by incorporating both fluid
and geometric parameters, offering insights into how mass
distribution affects spacetime curvature. Through orthogonal
decomposition of the Riemann tensor, a set of scalar quantities is
obtained, referred to the structure scalars, which serve as
indicators of celestial complexity. One specific scalar is then
specified as the complexity factor, i.e., $\mathbb{Y}_{TF}$,
facilitating further analysis on its role in characterizing complex
systems. The presence of unknowns in gravitational equations
necessitates the imposition of constraints to facilitate their
solution. To address this, $\mathbb{Y}_{TF}=0$ alongside three
distinct conditions are employed which yield diverse stellar models.
A comprehensive graphical analysis is conducted using multiple
values of the Rastall and charge parameters. Notably, the findings
of this study align with those predicted by Einstein's theory. More
appealingly, the Rastall theory demonstrates its superiority in the
presence of charge under model 2 when it is compared with the
general theory of relativity.
\end{abstract}

\section{Introduction}

The prevailing consensus in cosmological studies is that our cosmos
is experiencing a rapid expansion phase. Recent observational
evidences, encompassing studies of large-scale objects \cite{1c},
type Ia supernovae \cite{1a,1b} and CMB variations \cite{1d,1e},
have consistently supported this phenomenon. In addressing the
enigma of cosmic acceleration, researchers have primarily focused on
two avenues: investigating the role of dark energy or revising the
foundational principles of Einstein's theory of relativity (GR). The
accelerated expansion of the universe is often credited to a
mysterious entity known as ``dark energy'', whose exact
characteristics remain elusive. To uncover its underlying
properties, scientists have developed various theoretical models,
including $\Lambda$ (names the cosmological constant), phantom
paradigms \cite{1f,1g}, the Chaplygin gas \cite{1i}, tachyon fields
\cite{1h}, and others. A crucial tool for analyzing cosmic expansion
and its long-term journey is the equation of state for dark energy,
which represents the ratio between pressure and density within this
enigmatic component.

The pursuit of modifying GR has led to the creation of various
modified theories. Notably, Peter Rastall introduced a new theory in
1972 when he challenged conventional assumptions by suggesting that
the energy-momentum tensor (EMT), which typically has null
divergence in Euclidean geometries, could exhibit non-null
divergence in non-Euclidean or curved spacetimes \cite{1aa}. Such a
modification introduces a non-conserved EMT, where the Ricci scalar
plays a crucial role through an additional term known as the Rastall
parameter. This approach diverges from GR by incorporating a
coupling between matter and geometry that is beyond a minimal
interaction. The evolution of this theoretical framework has
significantly altered our understanding of how matter fields
interact with gravity, leading to a profound interplay between
matter distribution and geometric structures. By adopting this
framework, researchers have explored various phenomena and found
that it stands on par with other extensions of GR, which are derived
from modifications to the Einstein-Hilbert action
\cite{2aa}-\cite{7aaa}. A notable feature of Rastall theory is its
ability to provide solutions for perfect fluid interior geometries
that remain valid within the context of GR. Furthermore, in black
hole configurations, both GR and Rastall theory yield identical
results in vacuum regions.

The gravitational equations play a pivotal role in comprehensively
defining the architecture of celestial bodies. However, solving them
poses significant challenges, largely because of the intricate
involvement of geometric potentials and their derivatives,
especially when considering interactions between geometry and
matter. Solutions can be obtained through either analytical
techniques or numerical methods; the latter requires carefully
selected initial and boundary conditions tailored to the specific
context being studied. To develop a complete solution, additional
insights into local physical characteristics are required. Recent
studies on compact objects have underscored the potential impact of
diverse factors on the internal properties of celestial bodies,
particularly when considering perfect/imperfect fluid configurations
\cite{1}-\cite{3}. Furthermore, key physical components like density
variations, shear forces, and dissipation fluxes significantly
influence the stability of perfect fluid pressure condition in these
systems \cite{4}.

In the context of anisotropic spherical objects, the governing
equations involve three independent components with five unknowns,
comprising metric functions and a triplet associated with matter
variables. To obtain a unique solution, it is necessary to introduce
two additional conditions. These conditions often take the form of
empirical assumptions about physical parameters or equations of
state that relate various physical quantities \cite{5,6}. One common
approach is to utilize a polytropic equation \cite{7,8}, which has
been widely employed in studies examining the properties of
astrophysical white dwarfs-like bodies \cite{9}. The analysis has
been expanded to explore the systems admitting anisotropic
properties in GR, incorporating diverse theoretical approaches
\cite{10}-\cite{14}. Concurrently, constraints on geometric
functions have been constituted, exemplified by the Karmarkar
criteria, which involves selecting any one of these functions to
obtain the other \cite{15}-\cite{20}. Another viable option is
adopting the Weyl tensor set to zero (i.e., conformally flat
geometry) \cite{21}. These exhaustive analyzes demonstrate that the
structural evolution of massive systems can be conceptualized
through multiple frameworks, each governed by distinct conditions.

The study of complexity in dense systems has emerged as a
captivating and pivotal area of research. This inquiry is motivated
by the need to comprehend the characteristics of celestial bodies,
where forces due to gravity dominate particle interactions. 
The complexity observed in massive bodies is influenced by several
terms, including chaos dynamics that can lead to significant
structural changes even with minor alterations in key parameters,
resulting in unforeseen outcomes. The notion of complexity in fluid
dynamics underscores the intricate interactions among diverse
components within a fluid system, leading to emergent behaviors.
These components may include variations in density, directional
pressure distributions, energy transfer mechanisms, and other
relevant factors. Over time, researchers have sought to develop a
comprehensive definition of complexity that transcends disciplinary
boundaries across science. Initially, complexity was often framed in
terms of the information content and entropy within a system
\cite{22}-\cite{24}. However, when applied to contrasting systems
like perfect crystals and ideal gases - each with distinct
characteristics - it became apparent that this definition had
limitations. Despite their simplicity compared to complex systems,
these examples highlighted the need for a more nuanced understanding
of complexity.

A novel and influential complexity's definition, proposed by Herrera
\cite{25}, links this concept to specific physical parameters such
as variations in energy density and pressure anisotropy. Building on
Bel's method of the curvature tensor's orthogonal decomposition,
Herrera identified a set of scalars \cite{26,27}. Notably, he
encapsulated the critical factors like energy density inhomogeneity
and pressure anisotropy within a single scalar function, denoted by
$\mathbb{Y}_{TF}$, referred to the complexity factor. This approach
provides a nuanced framework for understanding complex systems by
quantifying their departure from idealized states. The concept
initially rested on the premise that complexity in a compact body
diminishes under two conditions: \emph{i} the internal geometry
exhibits homogeneous and isotropic characteristics, or \emph{ii} the
anisotropy and inhomogeneous energy density offset one another. This
framework has been expanded to explore evolutionary patterns in
objects with non-static and axial geometries, utilizing a complexity
factor originally proposed for static spherical structures
\cite{28,28a}.

Sharif and Naseer extended this notion by exploring charged
spherical systems with static interiors within the framework of GR
as well as a minimally coupled modified scenario. Their analysis
highlighted that $\mathbb{Y}_{TF}$ serves as a complexity factor in
both contexts \cite{30,31}. In a related study, Abbas and Nazar
\cite{32} developed novel models in $f(R)$ theory, investigating how
modifications influence both the mass and complexity of a compact
structure. The investigation into compact stars by them employed a
null complexity framework and a particular ansatz, yielding
promising outcomes in $f(R,T)$ framework \cite{33}. Meanwhile,
Manzoor and Shahid \cite{34} explored the evolutionary dynamics of a
pulsar $4U 1820-30$ using the Starobinsky model. Their findings
revealed that under specific parameter conditions, dark matter
density surpasses that of the conventional fluid by more than a
single order of magnitude.

This research builds upon previous investigations into anisotropic
models \cite{40} and explores deeper into the intricacies of
complexity within the framework of Rastall gravity. The organization
of this paper is as follows: Section \textbf{2} provides an overview
of the foundational principles underlying this non-conserved theory,
along with a detailed examination of the anisotropic EMT.
Subsequently, we derive and discuss the gravitational equations and
mass for a static sphere. Additionally, we assess the matching
conditions at the boundary to evaluate some parameters. In section
\textbf{3}, we delve into the intricacies of structure scalars,
highlighting $\mathbb{Y}_{TF}$ as a pivotal complexity factor.
Moving forward, section \textbf{4} offers a succinct examination of
the essential prerequisites that must be met to ensure a stellar
model's physical consistency. In section \textbf{5}, we offer a
triplet of celestial solutions, each subjected to graphical
interpretation under specific parametric settings. The final section
synthesizes our findings and juxtaposes them with other prominent
theories of gravity.

\section{Rastall Gravity and Electromagnetic Field}

Rastall theory diverges from the traditional notion that the
divergence of EMT is zero in flat spacetime, instead proposing that
it does not vanish in curved geometries \cite{1aa}. This framework
introduces novel field equations, which are derived from a modified
understanding of gravitational interactions under an electric charge
as
\begin{equation}\label{g1}
G_{\varphi\eta}\equiv
R_{\varphi\eta}-\frac{1}{2}Rg_{\varphi\eta}=\kappa
\big(T_{\varphi\eta}+E_{\varphi\eta}-\alpha Rg_{\varphi\eta}\big),
\end{equation}
where
\begin{itemize}
\item $G_{\varphi\eta}$ explains the geometric configuration and named the Einstein tensor and
$R_{\varphi\eta}$ being the Ricci tensor,
\item $\kappa$ refers to the coupling constant which shall be taken as unity,
\item $T_{\varphi\eta}$ being the EMT representing interior fluid
distribution,
\item $E_{\varphi\eta}$ being the EMT corresponding to an
electromagnetic field,
\item $\alpha$ symbolizes the Rastall parameter whose involvement preserves the fluid-geometry coupling.
\end{itemize}
A crucial insight emerges when these field equations align with the
non-conservation condition represented by
\begin{equation}\label{g2}
\nabla^\varphi (T_{\varphi\eta}+E_{\varphi\eta})=\alpha
g_{\varphi\eta}\nabla^\varphi R.
\end{equation}
Upon substituting $\alpha = 0$, the equation reduces to a form
consistent with GR. The presence of a non-null divergence in the
above equation is noteworthy, as it plays a pivotal role in
establishing the fluid-geometry interaction. By reinterpreting the
matter part of Eq.\eqref{g1}, we can express it as follows
\begin{equation}\label{g3}
\tilde{T}_{\varphi\eta}=T_{\varphi\eta}+E_{\varphi\eta}-\alpha
Rg_{\varphi\eta},
\end{equation}
making Eq.\eqref{g1} in a concise form as
\begin{eqnarray}\label{g4}
G_{\varphi\eta}=\kappa \tilde{T}_{\varphi\eta}.
\end{eqnarray}
Notably, adhering to the field equations outlined above yields the
standard conservation equation $\nabla^\varphi
\tilde{T}_{\varphi\eta} = 0$. This methodological framework is
readily applicable in various theoretical frameworks, where general
functions incorporating both matter and geometric terms supplant the
scalar curvature in the GR's action.

At this juncture in our investigation, our objective is to introduce
modifications to the Einstein field equations. By examining the
trace of Eq.\eqref{g1}, we can derive further insights into how
these modifications impact our understanding of gravitational
dynamics. The trace becomes
\begin{eqnarray}\label{g4a}
R=\frac{\kappa T}{4\alpha\kappa-1},
\end{eqnarray}
whose back substitution into Eq.\eqref{g3} yields
\begin{equation}\label{g4b}
\tilde{T}_{\varphi\eta}=T_{\varphi\eta}+E_{\varphi\eta}-\frac{\zeta
T}{4\zeta-1} g_{\varphi\eta},
\end{equation}
where $\zeta=\alpha\kappa$ which further becomes $\zeta=\alpha$ as
the coupling constant is taken to be the unity. It is necessary to
state the conditions under which one can achieve physically
realistic models in this framework, i.e., $\zeta \neq \frac{1}{4}$.

The exploration of Rastall theory is multifaceted, encompassing
diverse viewpoints. While certain researchers, like Visser
\cite{8aa} and Golovnev \cite{baa}, proposed that this gravity
essentially reformulates the EMT within the framework of GR, others
such as Darabi \emph{et al.} \cite{caa} countered their arguments by
suggesting that it involves a non-minimal coupling mechanism,
thereby distinguishing itself from GR. In contrast to Visser's
perspective, which suggested that the Rastall EMT can be used to
reconstruct the physical EMT, we propose a different interpretation.
If the goal is to demonstrate equivalence between Rastall and
Einstein theories, similar arguments could be applied to other
theories that couple matter and geometry, such as $f(R,T)$ gravity.
Moreover, while Visser argued for equivalence based on certain
mathematical transformations, our stance emphasizes that such
equivalences might not hold universally across all theoretical
frameworks. The non-conserved nature of the EMT in Rastall gravity
introduces unique dynamics compared to GR's conserved EMT.
Therefore, asserting that these two theories are identical overlooks
potential differences in their predictive power regarding
cosmological phenomena or quantum gravity challenges.

Anisotropy plays a pivotal role in shaping the behavior of compact
stellar objects by influencing their pressure profiles. In dense
systems, like neutron stars, the distinction between radial and
tangential pressures becomes critical due to the overwhelming
effects of gravity and magnetism on their internal structure. The
exploration of anisotropic fluid systems has seen significant
advancements since the early 20$^{th}$ century, with pivotal
research emphasizing the integration of anisotropic pressure into
stellar analyzes. For instance, numerous investigations have
established a foundation for understanding how anisotropic pressures
evolve, influenced by several factors \cite{37n}-\cite{37r}.
Following EMT represents such fluid content
\begin{equation}\label{g5}
T_{\varphi\eta}=(\rho+p_t)v_{\varphi}v_{\eta}+p_t
g_{\varphi\eta}+\left(p_r-p_t\right)u_\varphi u_\eta,
\end{equation}
where different physical terms such as radial pressure, tangential
pressure and energy density are symbolized by $p_r$, $p_t$ and
$\rho$, respectively. Further, $u_{\varphi}$ names the four-vector
and $v_{\varphi}$ refers to the four-velocity.

The electromagnetic field plays a crucial role in the structure and
dynamics of stars. It governs the interaction of charged particles,
influencing processes like energy transfer and magnetic field
generation within stellar interiors. These fields affect phenomena
such as stellar winds, magnetic storms, and sunspots. Additionally,
the electromagnetic forces contribute to the confinement of plasma
in stars, enabling the fusion reactions that power them. The EMT
representing the presence of charge in the considered fluid setup is
expressed as
\begin{equation}\label{g5a}
E_{\varphi\eta}=\frac{1}{4\pi}\left[\frac{1}{4}g_{\varphi\eta}F^{\tau\gamma}F_{\tau\gamma}
-F^{\gamma}_{\varphi}F_{\gamma\eta}\right],
\end{equation}
where the Maxwell field tensor is indicated by $F_{\varphi\eta}$
which is defined in terms of the four-potential
$\psi_{\eta}=\psi(r)\delta^{0}_{\eta}$ as
$F_{\varphi\eta}=\psi_{\eta;\varphi}-\psi_{\varphi;\eta}$.

The line element is fundamental as it defines the geometry of
spacetime, allowing us to measure distances, time intervals, and the
curvature of space. It is crucial for understanding the
gravitational interactions between objects, particularly in strong
fields such as black holes or neutron stars. By using the metric, we
can model the effects of gravity on light, matter, and time, helping
to describe the structure and behavior of the universe at both large
and small scales. In the current scenario, we assume a static sphere
given by
\begin{equation}\label{g6}
ds^2=-e^{\lambda_1(r)} dt^2+e^{\lambda_2(r)}
dr^2+r^2\big(d\theta^2+\sin^2\theta d\phi^2\big),
\end{equation}
under which the above four-quantities defined in Eq.\eqref{g5} takes
the values
\begin{equation}\label{g7}
u^\varphi=\delta^\varphi_1e^{\frac{-\lambda_2}{2}}, \quad
v^\varphi=\delta^\varphi_0e^{\frac{-\lambda_1}{2}},
\end{equation}
agreeing with the relations $u^\varphi v_{\varphi} = 0$, $u^\varphi
u_{\varphi} = 1$, and $v^\varphi v_{\varphi} = -1$.

The Maxwell equations $F^{\varphi\eta}_{;\eta}=4\pi
\jmath^{\varphi}$ under the metric \eqref{g6} produce
\begin{equation}\label{g5aa}
\psi''+\frac{1}{2r}\big\{4-r(\lambda_1'+\lambda_2')\big\}\psi'=4\pi\varrho
e^{\frac{\lambda_1}{2}+\lambda_2},
\end{equation}
where $'=\frac{d}{dr}$. Also, $\jmath^{\varphi}=\imath v^{\varphi}$
and $\imath$ are the current and charge densities, respectively. The
integration of Eq.\eqref{g5aa} gives
\begin{equation}
\psi'=\frac{q}{r^2}e^{\frac{\lambda_1+\lambda_2}{2}}, \quad q\equiv
q(r)=\int_0^r \imath e^{\frac{\lambda_2}{2}}\bar{r}^2d\bar{r},
\end{equation}
where the later term represents the total interior charge.

We now extract the non-vanishing components of Rastall's
gravitational equations utilizing Eqs.\eqref{g4}-\eqref{g5aa} as
\begin{align}\label{g8}
&e^{-\lambda_2}\left(\frac{\lambda_2'}{r}-\frac{1}{r^2}\right)
+\frac{1}{r^2}=\rho+\frac{q^2}{r^4}-\frac{\zeta}{4\zeta-1}\left(\rho-p_r-2p_t\right),\\\label{g9}
&e^{-\lambda_2}\left(\frac{1}{r^2}+\frac{\lambda_1'}{r}\right)
-\frac{1}{r^2}=p_r-\frac{q^2}{r^4}+\frac{\zeta}{4\zeta-1}\left(\rho-p_r-2p_t\right),
\\\label{g10}
&\frac{e^{-\lambda_2}}{4}\left[\lambda_1'^2-\lambda_1'\lambda_2'+2\lambda_1''-\frac{2\lambda_2'}{r}+\frac{2\lambda_1'}{r}\right]
=p_t+\frac{q^2}{r^4}+\frac{\zeta}{4\zeta-1}\left(\rho-p_r-2p_t\right),
\end{align}
which are further used to find explicit expressions of governing
physical parameters as
\begin{align}\nonumber
\rho&=\frac{e^{-\lambda_2}}{2r^4}\big[r^2 \big\{4 \zeta -r \lambda_2
' \big(4 \zeta +\zeta  r \lambda_1 '-2\big)+\zeta  r \big(2 r
\lambda_1 ''+\lambda_1 ' \big(r \lambda_1 '+4\big)\big)\\\label{g8a}
&-2\big\}-2 e^{\lambda_2 } \big\{q^2 +(2 \zeta -1)
r^2\big\}\big],\\\nonumber p_r&=\frac{e^{-\lambda_2}}{2r^4}\big[2
q^2 e^{\lambda_2 }+r^2 \big\{r \big(\zeta  \big(r \lambda_1 '+4\big)
\big(\lambda_2 '-\lambda_1 '\big)-2 \zeta  r \lambda_1 ''+2
\lambda_1 '\big)\\\label{g9a} &+2 (2 \zeta -1) \big(e^{\lambda_2
}-1\big)\big\}\big],\\\nonumber
p_t&=\frac{e^{-\lambda_2}}{4r^4}\big[r^2 \big\{r \big(\big(8 \zeta
+(2 \zeta -1) r \lambda_1 '-2\big) \big(\lambda_2 '-\lambda_1
'\big)+2 (1-2 \zeta ) r \lambda_1 ''\big)\\\label{g10a} &+8 \zeta
\big(e^{\lambda_2 }-1\big)\big\}-4 q^2  e^{\lambda_2 }\big].
\end{align}

Upon expanding the non-conservation phenomenon using the line
element \eqref{g6}, we obtain an expression that holds within the
Jordan frame (an extensive discussion is given in \cite{40a}. This
is given as
\begin{align}\label{g12}
\frac{dp_r}{dr}-\frac{qq'}{r^4}+\frac{\lambda_1'}{2}\left(\rho+p_r\right)+\frac{2\Pi}{r}
=\frac{\zeta}{4\zeta-1}\big(p_r'+2p_t'-\rho'\big),
\end{align}
which is Tolman-Oppenheimer-Volkoff equation in its generalized form
for anisotropic fluids \cite{40b} and $\Pi=p_r-p_t$ being the
anisotropy. This equation serves as a crucial instrument for
examining the evolutionary changes within celestial systems. It is
evident that the Rastall equations of motion \eqref{g8}-\eqref{g10}
incorporate multiple unknowns $(\lambda_1, \lambda_2, q, \rho, p_r,
p_t)$. As a result, imposing limitations on this set becomes
necessary to derive a well-behaved unique solution.

The Misner-Sharp mass for a static sphere associated with the
electric charge has the form
\begin{align}\label{g12a}
m(r)=\frac{r}{2}\left(1-e^{-\lambda_2}+\frac{q^2}{r^2}\right),
\end{align}
or, in other form using Eq.\eqref{g8}, it becomes
\begin{align}\label{g12b}
m(r)&=\frac{1}{2}\int_0^r\left\{\rho-\frac{\zeta}{4\zeta-1}\left(\rho-p_r-2p_t\right)\right\}
\bar{r}^2d\bar{r}+\int_0^r\frac{qq'}{\bar{r}}d\bar{r},
\end{align}
where
$\int_0^r\frac{qq'}{\bar{r}}d\bar{r}=\frac{1}{2}\int_0^r\frac{q^2}{\bar{r}^2}d\bar{r}+\frac{q^2}{2r}$.
The factor $\lambda_1'$ appeared above in Eq.\eqref{g12} can be
determined using \eqref{g9} and \eqref{g12a} which is given by
\begin{align}\label{g12c}
\lambda_1'=\frac{1}{r\big(r^2-2mr+q^2\big)}
\bigg[\bigg\{p_r+\frac{\zeta}{4\zeta-1}\left(\rho-p_r-2p_t\right)\bigg\}r^4-2q^2+2mr\bigg],
\end{align}
whose back substitution into Eq.\eqref{g12} leads to the following
\begin{align}\nonumber
&\frac{dp_r}{dr}-\frac{qq'}{r^4}+\frac{2\Pi}{r}+\frac{\zeta}{4\zeta-1}\big(\rho'-p_r'-2p_t'\big)
+\frac{\rho+p_r}{2r\big(r^2-2mr+q^2\big)}\\\label{g12d} &\times
\bigg[\bigg\{p_r+\frac{\zeta}{4\zeta-1}
\left(\rho-p_r-2p_t\right)\bigg\}r^4-2q^2+2mr\bigg]=0.
\end{align}

In order to ensure a smooth matching between the two (interior and
exterior) metrics at the boundary, junction conditions are utilized.
This approach facilitates a more detailed understanding of
structural evolution by providing a cohesive framework for the
current analysis. The external geometry is characterized by the
Reissner-Nordstr\"{o}m metric, which incorporates the total mass
$\texttt{M}$ and charge $\texttt{Q}$ as follows
\begin{equation}\label{g15}
ds^2=-\bigg(1-\frac{2\texttt{M}}{r}+\frac{\texttt{Q}^2}{r^2}\bigg)dt^2
+\bigg(1-\frac{2\texttt{M}}{r}+\frac{\texttt{Q}^2}{r^2}\bigg)^{-1}dr^2
+r^2\big(d\theta^2+\sin^2\theta d\phi^2\big).
\end{equation}
The matching of these metrics is predicated on satisfying specific
connections at $r=r_\Sigma=\texttt{R}$. The essential forms of
junction criteria are illustrated as
\begin{align}\label{g16}
e^{\lambda_1}~{_=^\Sigma}~1-\frac{2\texttt{M}}{\texttt{R}}
+\frac{\texttt{Q}^2}{\texttt{R}^2}~{_=^\Sigma}~e^{-\lambda_2}, \quad
p_r~{_=^\Sigma}~0.
\end{align}
The above equation (left) represents the alignment of both $g_{tt}$
and $g_{rr}$ components of two (inner and outer) spacetimes at the
surface, while the remaining equation ensures that radial pressure
vanishes at $r=\texttt{R}$.

\section{Adopting Complexity Factor among Structure Scalars under Orthogonal Splitting}

The concept of complexity in celestial systems has emerged as a
pivotal area of investigation in astrophysics. Various definitions
have been proposed in this regard, with one approach positing that
homogeneous and isotropic configurations exhibit minimal complexity.
This notion is grounded in the idea that irregularities in energy
density and anisotropies in the principal pressure contribute to the
overall complexity of such systems. A significant contribution to
this field was made by Herrera \cite{25}, who introduced a method
for quantifying complexity using specific scalar factors derived by
splitting the curvature tensor orthogonally \cite{26,27}. This
approach highlights how variations in energy density and pressure
influence the gravitational properties of these systems. In the
subsequent lines, we will provide an overview of the methodology
used to calculate the complexity factor, highlighting its
significance in relation to the aforementioned physical parameters.
The mathematical expression that demonstrates the decomposition of
$R^{\xi\varphi}_{\beta\vartheta}$ is formulated in terms of the Weyl
tensor $C^{\xi\varphi}_{\beta\vartheta}$ and other quantities as
follows
\begin{equation}\label{g23}
R^{\xi\varphi}_{\beta\vartheta}=C^{\xi\varphi}_{\beta\vartheta}+2
T^{[\xi}_{[\beta}\delta^{\varphi]}_{\vartheta]}+
T\left(\frac{1}{3}\delta^{\xi}_{[\beta}\delta^{\varphi}_{\vartheta]}
-\delta^{[\xi}_{[\beta}\delta^{\varphi]}_{\vartheta]}\right).
\end{equation}
The two tensors defined in terms of the Riemann tensor are given as
follows
\begin{eqnarray}\label{g24}
\mathbb{Y}_{\xi\beta}&=&R_{\xi\varphi\beta\vartheta}v^{\varphi}v^{\vartheta},\\\label{g25}
\mathbb{X}_{\xi\beta}&=&^{\ast}R^{\ast}_{\xi\varphi\beta\vartheta}v^{\varphi}v^{\vartheta}
=\frac{1}{2}\eta^{\omega\sigma}_{\xi\varphi}R^{\ast}_{\omega\sigma\beta\vartheta}
v^{\varphi}v^{\vartheta},
\end{eqnarray}
where $R^{\ast}_{\xi\varphi\beta\vartheta} = \frac{1}{2}
\eta_{\omega\sigma\beta\vartheta} R^{\omega\sigma}_{\xi\varphi}$ and
the Levi-Civita symbol is represented by
$\eta^{\omega\sigma}_{\xi\varphi}$. Another notation in which the
tensors \eqref{g24} and \eqref{g25} can be expressed in terms of the
projection tensor $h_{\xi\beta} = u_{\xi} u_{\beta} + g_{\xi\beta}$
and four-velocity as
\begin{eqnarray}\label{g26}
\mathbb{Y}_{\xi\beta}&=&\frac{1}{3}\big\{h_{\xi\beta}\mathbb{Y}_{T}+\big(3v_{\xi}v_{\beta}
-h_{\xi\beta}\big)\mathbb{Y}_{TF}\big\},\\\label{g27}
\mathbb{X}_{\xi\beta}&=&\frac{1}{3}\big\{h_{\xi\beta}\mathbb{X}_{T}+\big(3v_{\xi}v_{\beta}
-h_{\xi\beta}\big)\mathbb{X}_{TF}\big\}.
\end{eqnarray}
Through a series of straightforward yet lengthy computations (not
detailed in this work) utilizing Eqs.\eqref{g23}-\eqref{g27}, we
derive four scalars represented as
\begin{eqnarray}\label{g28}
&&\mathbb{X}_{T}=\rho+\frac{q^2}{r^4},\\\label{g28a}
&&\mathbb{X}_{TF}=-\mathbb{E}-\frac{\Pi}{2}+\frac{q^2}{r^4},\\\label{g28b}
&&\mathbb{Y}_{T}=\frac{1}{2}\big(\rho+3p_r-2\Pi\big)+\frac{q^2}{r^4},\\\label{g28c}
&&\mathbb{Y}_{TF}=\mathbb{E}-\frac{\Pi}{2}+\frac{q^2}{r^4}.
\end{eqnarray}
The only term which is newly introduced in these scalars is the
electric part of the Weyl tensor $\mathbb{E}$ whose value is defined
as
\begin{equation}\label{g29}
\mathbb{E}=\frac{e^{-\lambda_2}}{4}\left[\lambda_1''+\frac{\lambda_1'^2-\lambda_2'\lambda_1'}{2}
-\frac{\lambda_1'-\lambda_2'}{r}+\frac{2(1-e^{\lambda_2})}{r^2}\right].
\end{equation}

Our analysis reveals a strong connection between these scalar
functions and the intrinsic physical characteristics of celestial
interiors. By scrutinizing Eqs.\eqref{g28}-\eqref{g28c}, we gain
deeper insights into the evolution of self-gravitating structures,
as elaborated upon below
\begin{itemize}
\item The term $\mathbb{X}_{T}$ elucidates how homogeneous energy density influences fluid setup,

\item $\mathbb{X}_{TF}$ quantifies the extent of inhomogeneous energy density,

\item $\mathbb{Y}_{T}$ modulates the anisotropy at local scale,

\item $\mathbb{Y}_{TF}$ serves as a function, encompassing both the roles of
$\mathbb{X}_{TF}$ and $\mathbb{Y}_{T}$.
\end{itemize}

To highlight the intricate relationships among these elements,
specific modifications are required. This involves linking the mass
function, Weyl scalar and fluid parameters in a manner that reveals
their interconnectedness. This relation is given as
\begin{align}\nonumber
m&=\frac{r^3}{6}\big(\rho-p_r+p_t\big)
-\frac{r^3\mathbb{E}}{3}+\frac{q^2}{r}+\frac{1}{3} r e^{-\lambda_2 }
\big(e^{\lambda_2 }-1\big)-\frac{1}{24} r e^{-\lambda_2 }\big[r
\big\{4 \zeta  r \lambda_1 ''\\\label{g29a} &+\big(2 (4 \zeta -1)+2
\zeta r \lambda_1 '+2\big) \big(\lambda_1 '-\lambda_2 '\big)\big\}-8
(\zeta -1) \big(e^{\lambda_2 }-1\big)\big].
\end{align}
When the above equation is combined with \eqref{g12b}, this results
in
\begin{align}\nonumber
\mathbb{E}&=\frac{e^{-\lambda_2}}{4 r^4}\big[\zeta  r^2 \big\{r
\big(\big(r \lambda_1 '+4\big) \big(\lambda_2 '-\lambda_1 '\big)-2 r
\lambda_1 ''\big)-4\big\}\\\label{g29b} &-2 e^{ \lambda_2 } \big\{r
\big(6 m+r^3 (\Pi -\rho )-2 \zeta r\big)-6 q^2\big\}\big],
\end{align}
Substituting this into Eq.\eqref{g28c} makes sure that the scalar
$\mathbb{Y}_{TF}$ possess all those factors which govern the
complexity. The following expression for $\mathbb{Y}_{TF}$ is
provided as an evident
\begin{align}\nonumber
\mathbb{Y}_{TF}&=\frac{e^{-\lambda_2}}{4 r^4}\big[\zeta  r^2 \big\{r
\big(\big(r \lambda_1 '+4\big) \big(\lambda_2 '-\lambda_1 '\big)-2 r
\lambda_1 ''\big)-4\big\}-2 e^{\lambda_2 }\\\label{g30}
&\times\big\{r \big(6 m+r^3 (\Pi -\rho )-2 \zeta r\big)-6
q^2\big\}\big]-\frac{\Pi }{2}+\frac{q^2}{r^4}.
\end{align}
It is crucial to note that this scalar vanishes when the fluid
composition is considered perfect and homogeneous. Additionally, a
newly developed Tolman mass enhances our ability to estimate the
total mass-energy of a structure, yet failed to ensure precise
localization \cite{aa}.

It is crucial to emphasize that a configuration devoid of complexity
cannot be solely attributed to a homogeneous and isotropic
configuration. Instead, this condition can be fulfilled when
$\mathbb{Y}_{TF}$ equals zero. Using this criterion with
Eq.\eqref{g30}, we derive an expression that interconnects the
fundamental fluid parameters, thereby establishing their mutual
relationships as
\begin{equation}\label{g34}
\Pi=\frac{e^{-\lambda_2}}{4r^4}\big[2 e^{\lambda_2 } \big\{5 q^2+r^2
\big(2 \zeta +\rho r^2-3\big)\big\}+r^2 \big\{\zeta  r \big(\big(r
\lambda_1 '+4\big)\big(\lambda_2 '-\lambda_1 '\big)-2 r \lambda_1
''\big)+6-4 \zeta\big\}\bigg],
\end{equation}
following that there is a distinct category of solutions fulfilling
this criterion, as noted in earlier studies \cite{40}. Since this
criterion corresponds to a non-local equation of state \cite{ai}, it
enhances our ability to address the governing equations effectively.
This study aims to explore the implications of anisotropic EMTs by
examining scenarios where they naturally occur, such as in
environments characterized by unusual matter distributions or near
extremely dense astrophysical objects. Investigating these contexts
can provide crucial insights into the significance and impact of
this condition in real astrophysical settings. Consequently,
neglecting this aspect may lead to fundamentally different outcomes
in theoretical models, as suggested by previous research on similar
topics \cite{22}.

\section{Brief Review of Some Physical Requirements for Existence}

Researchers have developed multiple methodologies to address the
field equations that govern celestial bodies of physical
significance. However, when these solutions fail to satisfy
established standards, they are often considered unsuitable for
modeling realistic compact stars. Various conditions have been
proposed and upheld by different researchers in this context, which
are detailed below \cite{ab,ac}.
\begin{itemize}
\item In a self-gravitating fluid interior, it is essential for the metric
functions to remain bounded and singularity-free, ensuring they
remain positive.

\item The peak values of fluid triplet such as density and
principal pressures should be located in the core ($r = 0$),
indicating a consistent and positive trend across the entire domain.
At $r = 0$, their first derivatives should vanish while their second
derivatives should be negative, reflecting a diminishing pattern as
they approach the boundary.

\item The arrangement of particles within a dense structure influences
their mutual proximity, thereby defining the system's compactness.
Additionally, the mass-to-radius ratio plays a crucial role in
characterizing such systems, particularly for static spheres where
this ratio must adhere to specific constraints in GR as outlined in
relevant studies
\begin{align}\nonumber
\texttt{R}-\sqrt{\texttt{R}^2-2\texttt{M}\texttt{R}+\texttt{Q}^2}
-\bigg(\texttt{M}-\frac{\texttt{Q}^2}{2\texttt{R}}\bigg)\leq
\frac{1}{2}\bigg(\texttt{M}-\frac{\texttt{Q}^2}{2\texttt{R}}\bigg),
\end{align}
where $\texttt{M}-\frac{\texttt{Q}^2}{2\texttt{R}}\neq0$. On the
other hand, the alternative definition of this factor takes the form
\begin{align}\nonumber
\frac{\texttt{M}}{\texttt{R}} \leq
\frac{8}{9}\bigg\{\frac{1}{1+\sqrt{1-\frac{8\Theta}{9}}}\bigg\},
\quad \Theta=\frac{\texttt{Q}^2}{\texttt{M}^2},
\end{align}
whose equivalent notation is described by \cite{42ga,42gb}
\begin{align}\label{g49}
\frac{1}{\texttt{R}}\bigg(2\texttt{M}-\frac{\texttt{Q}^2}{\texttt{R}}\bigg)
\leq \frac{8}{9}.
\end{align}
However, deviations arise when considering alternative gravity
theories or fluid distributions with pressure anisotropy. These
modifications stem from shifts in gravitational interactions and
variations in internal pressure behavior \cite{adada}-\cite{adadf}.
Anisotropic pressure plays a pivotal role in determining the
equilibrium and density limits of compact stars. Research indicates
that such anisotropy can revise the traditional compactness
constraints by introducing new force contributions that either
resist or amplify gravitational contraction. For instance, positive
anisotropy produces repulsive effects, bolstering stability and
permitting higher compactness ratios. Conversely, negative
anisotropy introduces compressive forces, potentially reducing the
allowable compactness. In Rastall gravity, the introduction of
non-minimal coupling modifies gravitational interactions,
influencing key stellar characteristics. This investigation focuses
on obtaining numerical solutions for the strongly nonlinear
differential equations that emerge in Rastall's framework. Due to
these complexities, exact analytical solutions for the adjusted
Buchdahl bound remain unattainable. Instead, numerical estimates for
this limit are evaluated for the parameter values, including $\zeta
= 0,~0.1,~0.2$. They are
\begin{itemize}
\item For $\chi=0$, these values are 0.889, 0.772, and 0.714, respectively \cite{sl5}.
\item For $\chi=0.2$, these values are 0.889, 0.771, and 0.711, respectively.
\end{itemize}
Notably, the Buchdahl limit in Rastall theory is marginally reduced
when compared to its counterpart in GR.

\item The existence of ordinary matter within a celestial body's core
is contingent upon fulfilling specific criteria. These criteria are
defined as energy bounds, which encompass various mergers of fluid
parameters inherent to the relevant EMT. In the context of our
analysis, these conditions involve
\begin{equation}
\left.
\begin{aligned}\label{g50}
&\rho+p_r \geq 0, \quad \rho+p_t+\frac{2q^2}{r^4} \geq 0, \\
&\rho-p_r+\frac{2q^2}{r^4} \geq 0, \quad \rho-p_t \geq 0,\\
&\rho+\frac{q^2}{r^4} \geq 0, \quad \rho+p_r+2p_t+\frac{2q^2}{r^4}
\geq 0.
\end{aligned}
\right\}
\end{equation}
Among all above, two conditions are most important to check when the
fluid parameters show positive behavior for all values of $r$,
referred to dominant energy bounds which are $\rho - p_{r} +
\frac{2q^2}{r^4} \geq 0$ and $\rho - p_{t} \geq 0$. This also
ensures that the pressure components do not exceed the fluid's
density.

\item In the context of stellar objects, gravitational redshift
is often expressed as $z = e^{-\lambda_1 / 2} - 1$. It is crucial
for this model's validity that $z$ decreases as the radial distance
increases. Moreover, ensuring that its value remains below a certain
threshold , i.e., $5.211$ at the boundary radius $r=\texttt{R}$ is
essential for maintaining consistency within theoretical frameworks
\cite{ad}.

\item Recent studies have focused on examining the structure's stability
that exhibit minor departure from the state of being in hydrostatic
equilibrium, a topic of considerable interest in contemporary
research. The concept of cracking, initially explored by Herrera
\emph{et al.} \cite{ae,af}, arises when the total force in radial
direction changes sign at a specific point due to external
influences. To prevent such cracking, it is essential that the
following condition on the difference between tangential $v_{t}^{2}
= \frac{dp_{t}}{d\rho}$ and radial $v_{r}^{2} =
\frac{dp_{r}}{d\rho}$ sound speed components holds true
\begin{align}\label{g51}
0 \leq v_{r}^{2}-v_{t}^{2} \leq 1.
\end{align}
\end{itemize}

\section{Formulation of Three Stellar Models via A Novel Approach}

Numerous models have been developed by researchers to address the
field equations. For instance, Herrera's \cite{25} work involved
employing two specific constraints to obtain solutions. In the
current analysis, we apply various conditions to derive multiple
solutions. These solutions shall then be evaluated based on their
physical characteristics through graphical representations for a
range of parameter values.

\subsection{$\mathbb{Y}_{TF}=0=p_r$ Constraints}

To address the field equations \eqref{g8a}-\eqref{g10a}, which
encompass six unknown variables ($\rho$, $p_r$, $p_t$, $q$,
$\lambda_1$, $\lambda_2$), supplementary constraints are required.
In order to simplify the problem, we adopt the assumptions
$\mathbb{Y}_{TF} = 0$ and $p_r = 0$ along with known electric charge
as $q=\sqrt{\chi}r^3$ with $\chi$ being a constant \cite{aha},
resulting in a unique solution that solely includes the tangential
pressure component. This approach aligns with the notion established
by Florides \cite{ah}. When combining the later constraint with
Eq.\eqref{g9a}, it yield a differential equation given as follows
\begin{align}\label{g52}
2 \chi r^6 e^{\lambda_2 }+r^2 \big[r \big\{\zeta  \big(r \lambda_1
'+4\big) \big(\lambda_2 '-\lambda_1 '\big)-2 \zeta  r \lambda_1 ''+2
\lambda_1 '\big\}+2 (2 \zeta -1) \big(e^{\lambda_2 }-1\big)\big]=0,
\end{align}
whereas joining Eqs.\eqref{g8a}, \eqref{g10a}, and \eqref{g34}, the
former constraint results in
\begin{align}\nonumber
&4 \big\{2 \zeta -e^{\lambda_2 } \big(2 \zeta +r^4 \chi
-1\big)-1\big\}+r \big[\big\{8 \zeta +\big(2 \zeta -1\big) r
\lambda_1 '\big\}\\\label{g53} &\times \big(\lambda_1 '-\lambda_2
'\big)+2 (2 \zeta -1) r \lambda_1 ''-2 \lambda_1 '\big]=0.
\end{align}

The inclusion of curvature terms in the modified framework results
in the previous equations becoming fourth-order with respect to the
spacetime potentials. Consequently, obtaining an exact analytical
solution is not feasible. To determine the values of $g_{rr}$ and
$g_{tt}$ components, we employ numerical integration technique by
utilizing carefully selected initial conditions. As illustrated in
Figure \textbf{1}, the curves of $e^{\lambda_1}$ and
$e^{-\lambda_2}$ display a consistently positive and
singularity-free behavior. At $r=0$, we get $e^{\lambda_1(0)}=c_1$
(with $c_1$ being positive constant) and $e^{-\lambda_2(0)} = 1$
align with our initial expectations. The boundary or outermost
extent of the star is actually the point of intersection of these
two functions. In the following, we provide the summary.
\begin{enumerate}
\item In this model, we extract the values of radius along with the
compactness for $\chi=0$ to further analyze its properties as
\begin{itemize}
\item For $\zeta = 0$, the value of $\texttt{R}$ is 0.12, and the
compactness is
$$e^{\lambda_1(0.12)}=e^{-\lambda_2(0.12)}\approx0.28=1-\frac{2\texttt{M}}{\texttt{R}}
+\frac{\texttt{Q}^2}{\texttt{R}^2} \quad \Rightarrow \quad
\frac{2\texttt{M}}{\texttt{R}}-\frac{\texttt{Q}^2}{\texttt{R}^2}
\approx0.72 < 0.889.$$
\item For $\zeta = 0.1$, the value of $\texttt{R}$ is 0.14, and the compactness
is
$$e^{\lambda_1(0.14)}=e^{-\lambda_2(0.14)}\approx0.36=1-\frac{2\texttt{M}}{\texttt{R}}
+\frac{\texttt{Q}^2}{\texttt{R}^2} \quad \Rightarrow \quad
\frac{2\texttt{M}}{\texttt{R}}-\frac{\texttt{Q}^2}{\texttt{R}^2}
\approx0.64 < 0.772.$$
\item For $\zeta = 0.2$, the value of $\texttt{R}$ is 0.19, and the compactness
is
$$e^{\lambda_1(0.19)}=e^{-\lambda_2(0.19)}\approx0.82=1-\frac{2\texttt{M}}{\texttt{R}}
+\frac{\texttt{Q}^2}{\texttt{R}^2} \quad \Rightarrow \quad
\frac{2\texttt{M}}{\texttt{R}}-\frac{\texttt{Q}^2}{\texttt{R}^2}
\approx0.18 < 0.714.$$
\end{itemize}

\item On the other hand, these values for $\chi=0.2$ are
\begin{itemize}
\item For $\zeta = 0$, the value of $\texttt{R}$ is 0.14, and the
compactness is
$$e^{\lambda_1(0.14)}=e^{-\lambda_2(0.14)}\approx0.18=1-\frac{2\texttt{M}}{\texttt{R}}
+\frac{\texttt{Q}^2}{\texttt{R}^2} \quad \Rightarrow \quad
\frac{2\texttt{M}}{\texttt{R}}-\frac{\texttt{Q}^2}{\texttt{R}^2}
\approx0.82 < 0.889.$$
\item For $\zeta = 0.1$, the value of $\texttt{R}$ is 0.16, and the compactness
is
$$e^{\lambda_1(0.16)}=e^{-\lambda_2(0.16)}\approx0.26=1-\frac{2\texttt{M}}{\texttt{R}}
+\frac{\texttt{Q}^2}{\texttt{R}^2} \quad \Rightarrow \quad
\frac{2\texttt{M}}{\texttt{R}}-\frac{\texttt{Q}^2}{\texttt{R}^2}
\approx0.74 < 0.771.$$
\item For $\zeta = 0.2$, the value of $\texttt{R}$ is 0.2, and the compactness
is
$$e^{\lambda_1(0.2)}=e^{-\lambda_2(0.2)}\approx0.8=1-\frac{2\texttt{M}}{\texttt{R}}
+\frac{\texttt{Q}^2}{\texttt{R}^2} \quad \Rightarrow \quad
\frac{2\texttt{M}}{\texttt{R}}-\frac{\texttt{Q}^2}{\texttt{R}^2}
\approx0.2 < 0.711.$$
\end{itemize}
\end{enumerate}
\begin{figure}\center
\epsfig{file=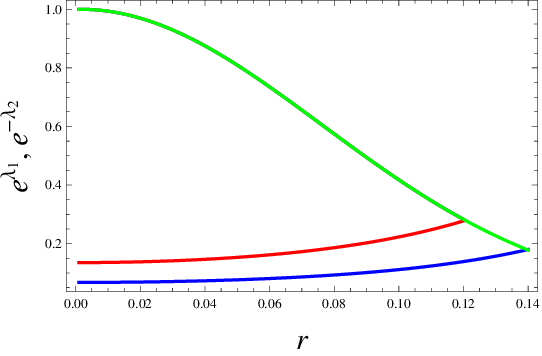,width=0.4\linewidth}\epsfig{file=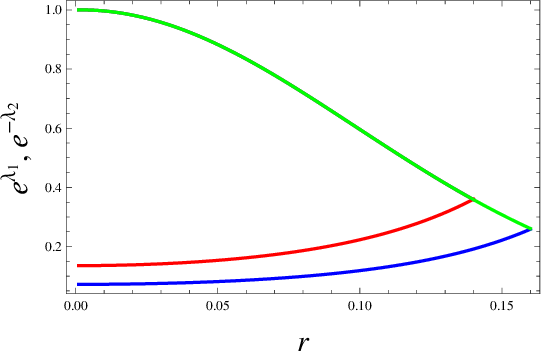,width=0.4\linewidth}
\epsfig{file=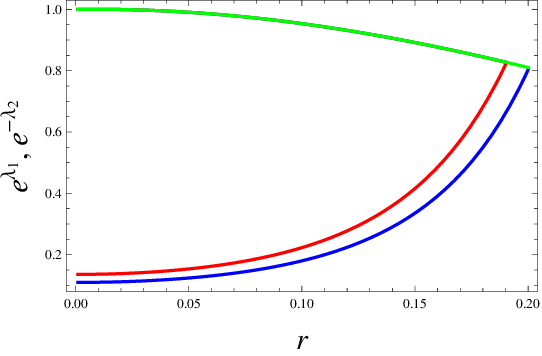,width=0.4\linewidth}
\caption{Geometric potentials $e^{\lambda_1}$
(\textcolor{red}{\textbf{\small $\bigstar$}}) ($\chi=0$) and
(\textcolor{blue}{\textbf{\small $\bigstar$}}) ($\chi=0.2$), and
$e^{-\lambda_2}$ (\textcolor{green}{\textbf{\small $\bigstar$}}) for
model 1 [$\zeta=0$ (left), $0.1$ (right) and $0.2$ (lower)].}
\end{figure}

Figure \textbf{2} highlights how the fluid's density takes a
distinct trend with its maximum at $r=0$, tapering off as it
approaches the outer surface. The influence of both Rastall
corrections and charge on the internal structure leads to lower
densities compared to those derived from GR \cite{40}. This scenario
aligns with solutions like Florides', where radial pressure is
negligible; thus, stability remains intact until tangential
pressures rise as radii increase. The profile of $p_t$ is also
depicted, aligning with the anticipated profile. Notably, the
anisotropy exhibits a divergent trend compared to the tangential
pressure, which is represented by $\Pi=-p_t$ (as shown in the same
Figure).

The viability analysis is illustrated in Figure \textbf{3},
exhibiting a positive trend, thereby supporting this developed
solution except for $\zeta=0$ along with with $\chi=0.2$. The left
panel of Figure \textbf{4} visualizes the gravitational redshift,
which takes less values as $r$ increases. Specifically, at the
boundary, its calculated values are $0.882$, $0.611$, and $0.084$
for $\zeta = 0$, $0.1$, and $0.2$, and $\chi=0$ respectively.
Similarly, we obtain these values as $z(0.14) \approx 0.234$,
$z(0.16) \approx 0.215$, and $z(0.20) \approx 0.001$ for $\chi=0.2$.
The observed values are significantly lower than the maximum
threshold reported by researchers, specifically $z({\texttt{R}}) =
5.211$. Furthermore, the stability analysis indicates that this
solution exhibits the sign of cracking only for $\zeta=0$ along with
$\chi=0.2$. Hence, the structure maintains its stability for all
other values.
\begin{figure}\center
\epsfig{file=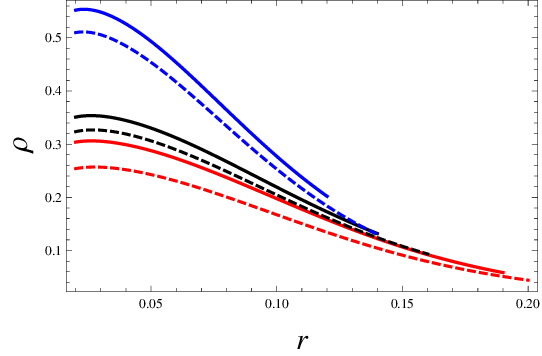,width=0.4\linewidth}\epsfig{file=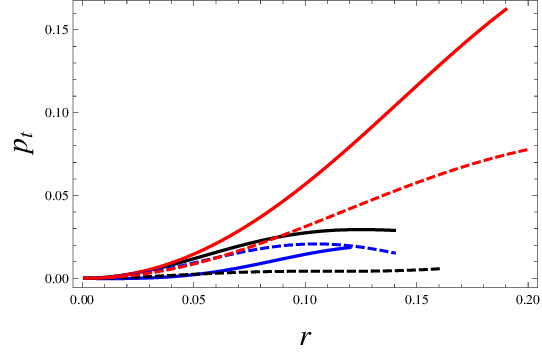,width=0.4\linewidth}
\epsfig{file=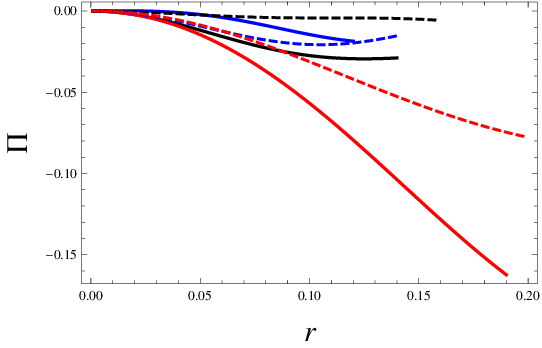,width=0.4\linewidth} \caption{Fluid
variables for $\zeta=0$ (\textcolor{blue}{\textbf{\small
$\bigstar$}}), $0.1$ (\textcolor{black}{\textbf{\small $\bigstar$}})
and $0.2$ (\textcolor{red}{\textbf{\small $\bigstar$}}) for model 1
[$\chi=0$ (solid) and $0.2$ (dashed)].}
\end{figure}
\begin{figure}\center
\epsfig{file=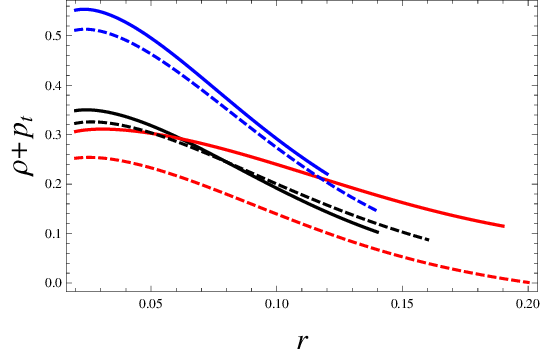,width=0.4\linewidth}\epsfig{file=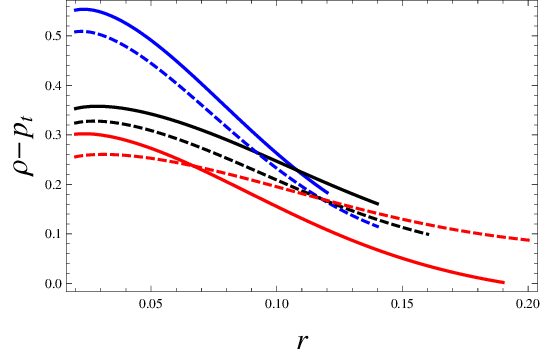,width=0.4\linewidth}
\epsfig{file=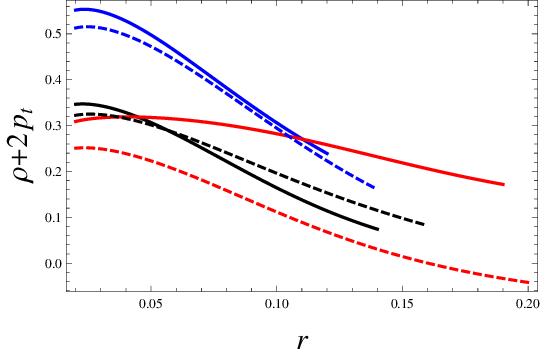,width=0.4\linewidth} \caption{Viability for
$\zeta=0$ (\textcolor{blue}{\textbf{\small $\bigstar$}}), $0.1$
(\textcolor{black}{\textbf{\small $\bigstar$}}) and $0.2$
(\textcolor{red}{\textbf{\small $\bigstar$}}) for model 1 [$\chi=0$
(solid) and $0.2$ (dashed)].}
\end{figure}
\begin{figure}\center
\epsfig{file=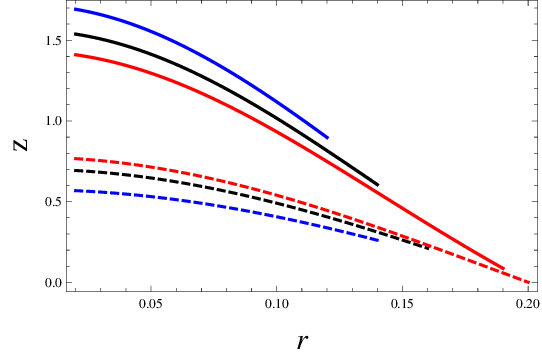,width=0.4\linewidth}\epsfig{file=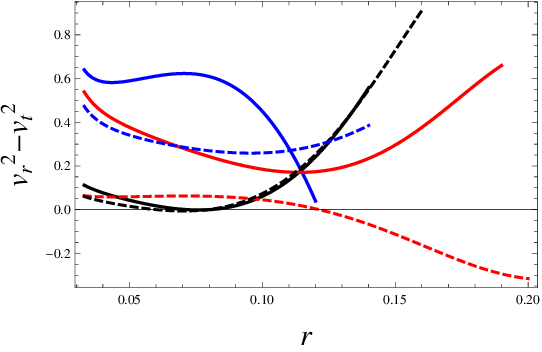,width=0.4\linewidth}
\caption{Redshift and stability for $\zeta=0$
(\textcolor{blue}{\textbf{\small $\bigstar$}}), $0.1$
(\textcolor{black}{\textbf{\small $\bigstar$}}) and $0.2$
(\textcolor{red}{\textbf{\small $\bigstar$}}) for model 1 [$\chi=0$
(solid) and $0.2$ (dashed)].}
\end{figure}

\subsection{$\mathbb{Y}_{TF}=0$ and
$p_r=\mathcal{I}\rho^{\eta_3}=\mathcal{I}\rho^{1+\frac{1}{\mathcal{U}}}$
Constraints}

In contemporary studies, the polytropic model associated with
anisotropic fluid plays a crucial role. Various researchers have
investigated polytropic solutions within diverse gravitational
frameworks, contributing significantly to our understanding of these
phenomena \cite{12}-\cite{14}. To solve the field equations
\eqref{g8a}-\eqref{g10a}, we adopt a polytropic equation and set the
complexity factor to zero, thereby simplifying our analysis. A
comprehensive overview of the polytropic model can be found in
\cite{25}, yet our approach delves deeper into the solution,
utilizing visual aids to enhance understanding. Building on the
conditions outlined previously, we proceed with our analysis as
follows
\begin{align}\label{g55}
p_r=\mathcal{I}\rho^{\eta_3}=\mathcal{I}\rho^{1+\frac{1}{\mathcal{U}}},
\quad \mathbb{Y}_{TF}=0,
\end{align}
where
\begin{itemize}
\item $\mathcal{U}$ indicates a polytropic index,
\item $\mathcal{I}$ symbolizes a constant,
\item $\eta_3$ represents a polytropic exponent.
\end{itemize}
We get two higher-order equations in metric components from the
constraints provided in Eq.\eqref{g55}. These equations take the
form
\begin{align}\nonumber
&\frac{e^{-\lambda_2}}{2r^4}\big[2 q^2 e^{\lambda_2 }+r^2 \big\{r
\big(\zeta  \big(r \lambda_1 '+4\big) \big(\lambda_2 '-\lambda_1
'\big)-2 \zeta  r \lambda_1 ''+2 \lambda_1 '\big)+2 (2 \zeta
-1)\\\nonumber &\times \big(e^{\lambda_2
}-1\big)\big\}\big]-\mathcal{I}\big(\frac{e^{-\lambda_2}}{2r^4}\big)^{1+\frac{1}{\mathcal{U}}}
\big[r^2 \big\{4 \zeta -r \lambda_2 ' \big(4 \zeta +\zeta  r
\lambda_1 '-2\big)+\zeta  r\big(2 r \lambda_1 '' \\\label{g55a}
&+\lambda_1 ' \big(r \lambda_1 '+4\big)\big)-2\big\}-2 e^{\lambda_2
} \big\{q^2 +(2 \zeta -1)
r^2\big\}\big]^{1+\frac{1}{\mathcal{U}}}=0,\\\label{g55b} &\lambda_1
' \big(r \lambda_2 '-r \lambda_1 '+2\big)-2 r \lambda_1 ''=0.
\end{align}
By setting $\mathcal{U} = 0.05$ and $\mathcal{I} = 0.9$, we proceed
to solve above equations, thereby obtaining the values of
$\lambda_1$ and $\lambda_2$. The numerical solution of these
equations allows us to visualize the graphical behavior of the
resulting metric potentials, as depicted in Figure \textbf{5}, which
exhibits a consistent pattern. Significantly, the explicit forms of
these variables cannot be directly obtained due to the necessity of
numerical solutions for the aforementioned equations. The results
indicate that $e^{\lambda_1(0)}$ equals a positive constant denoted
as $c_2$, while $e^{-\lambda_2(0)}$ evaluates to unity.
\begin{enumerate}
\item Across different values of $\zeta$, variations in both the
boundary and compactness are observed for $\chi=0$ as
\begin{itemize}
\item For $\zeta = 0$, the value of $\texttt{R}$ is 0.095, and the compactness
is
$$e^{\lambda_1(0.095)}=e^{-\lambda_2(0.095)}\approx0.21=1-\frac{2\texttt{M}}{\texttt{R}}
+\frac{\texttt{Q}^2}{\texttt{R}^2} \quad \Rightarrow \quad
\frac{2\texttt{M}}{\texttt{R}}-\frac{\texttt{Q}^2}{\texttt{R}^2}
\approx0.79 < 0.889.$$
\item For $\zeta = 0.1$, the value of $\texttt{R}$ is 0.10, and the compactness
is
$$e^{\lambda_1(0.10)}=e^{-\lambda_2(0.10)}\approx0.36=1-\frac{2\texttt{M}}{\texttt{R}}
+\frac{\texttt{Q}^2}{\texttt{R}^2} \quad \Rightarrow \quad
\frac{2\texttt{M}}{\texttt{R}}-\frac{\texttt{Q}^2}{\texttt{R}^2}
\approx0.64 < 0.772.$$
\item For $\zeta = 0.2$, the value of $\texttt{R}$ is 0.12, and the compactness
is
$$e^{\lambda_1(0.12)}=e^{-\lambda_2(0.12)}\approx0.54=1-\frac{2\texttt{M}}{\texttt{R}}
+\frac{\texttt{Q}^2}{\texttt{R}^2} \quad \Rightarrow \quad
\frac{2\texttt{M}}{\texttt{R}}-\frac{\texttt{Q}^2}{\texttt{R}^2}
\approx0.46 < 0.714.$$
\end{itemize}

\item On the other hand, these values for $\chi=0.2$ are
\begin{itemize}
\item For $\zeta = 0$, the value of $\texttt{R}$ is 0.11, and the compactness
is
$$e^{\lambda_1(0.11)}=e^{-\lambda_2(0.11)}\approx0.13=1-\frac{2\texttt{M}}{\texttt{R}}
+\frac{\texttt{Q}^2}{\texttt{R}^2} \quad \Rightarrow \quad
\frac{2\texttt{M}}{\texttt{R}}-\frac{\texttt{Q}^2}{\texttt{R}^2}
\approx0.87 < 0.889.$$
\item For $\zeta = 0.1$, the value of $\texttt{R}$ is 0.12, and the compactness
is
$$e^{\lambda_1(0.12)}=e^{-\lambda_2(0.12)}\approx0.24=1-\frac{2\texttt{M}}{\texttt{R}}
+\frac{\texttt{Q}^2}{\texttt{R}^2} \quad \Rightarrow \quad
\frac{2\texttt{M}}{\texttt{R}}-\frac{\texttt{Q}^2}{\texttt{R}^2}
\approx0.76 < 0.771.$$
\item For $\zeta = 0.2$, the value of $\texttt{R}$ is 0.14, and the compactness
is
$$e^{\lambda_1(0.14)}=e^{-\lambda_2(0.14)}\approx0.43=1-\frac{2\texttt{M}}{\texttt{R}}
+\frac{\texttt{Q}^2}{\texttt{R}^2} \quad \Rightarrow \quad
\frac{2\texttt{M}}{\texttt{R}}-\frac{\texttt{Q}^2}{\texttt{R}^2}
\approx0.57 < 0.711.$$
\end{itemize}
\end{enumerate}
\begin{figure}\center
\epsfig{file=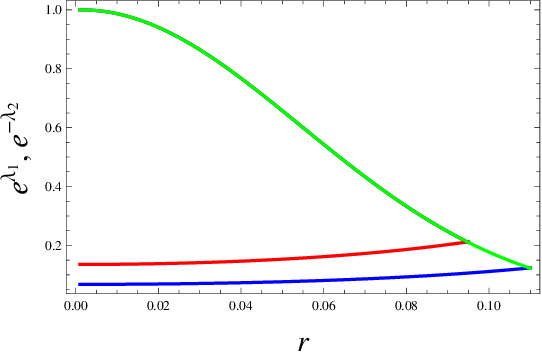,width=0.4\linewidth}\epsfig{file=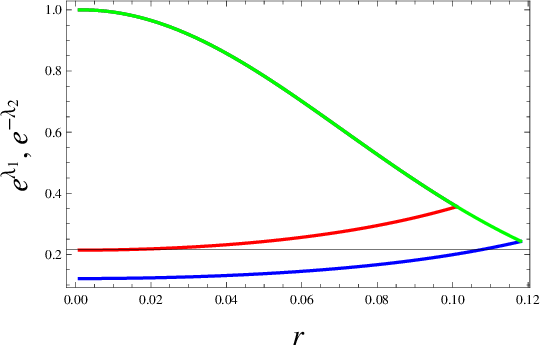,width=0.4\linewidth}
\epsfig{file=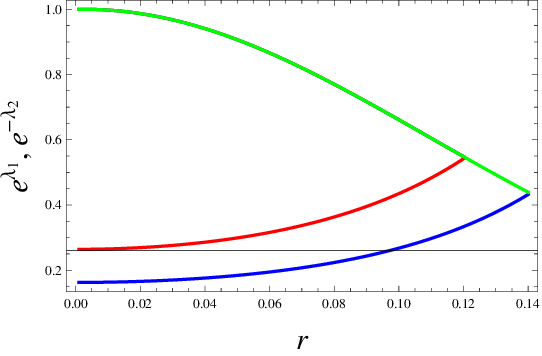,width=0.4\linewidth}
\caption{Geometric potentials $e^{\lambda_1}$
(\textcolor{red}{\textbf{\small $\bigstar$}}) ($\chi=0$) and
(\textcolor{blue}{\textbf{\small $\bigstar$}}) ($\chi=0.2$), and
$e^{-\lambda_2}$ (\textcolor{green}{\textbf{\small $\bigstar$}}) for
model 2 [$\zeta=0$ (left), $0.1$ (right) and $0.2$ (lower)].}
\end{figure}

Figure \textbf{6} presents the radial/tangential pressures' profiles
as well as the energy density of the developed fluid distribution.
These quantities reach their maximum values at the core and decrease
monotonically to their minimum values at the boundary. Both
parameters $\zeta$ and $\chi$ are found to be inversely proportional
to the energy density. The second plot reveals that $p_r$ vanishes
at the interface for all choices of $\zeta$ and $\chi$. Figure
\textbf{7} presents comprehensive plots of all viability conditions,
demonstrating satisfaction across every $\zeta$. Consequently, these
conditions collectively form a physically viable solution, implying
the presence of conventional matter.

In Figure \textbf{8}, the gravitational redshift is illustrated for
this solution, demonstrating a decrease as $r$ increases.
Specifically, at $r = 0.095$, $0.10$, and $0.12$, its values are
calculated as $1.159$, $0.996$, and $0.745$ for $\zeta$ set to $0$,
$0.1$, and $0.2$ and $\chi=0$, respectively. Similarly, these values
are obtained as $0.413$, $0.356$, and $0.243$ for $\chi=0.2$.
Furthermore, this Figure examines the stability region, revealing
that cracking phenomena occur exclusively when $\zeta = 0 = \chi$.
This observation highlights a critical threshold beyond which
structural integrity is compromised. The adoption of this value
leads to system's instability. Conversely, the other two values of
$\zeta$ along with both choices of charge parameter yield physically
stable interior. Consequently, Rastall theory provides more
advantageous results compared to those obtained from GR \cite{40}.
\begin{figure}\center
\epsfig{file=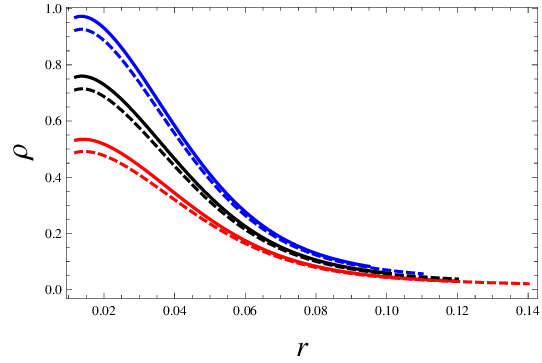,width=0.4\linewidth}\epsfig{file=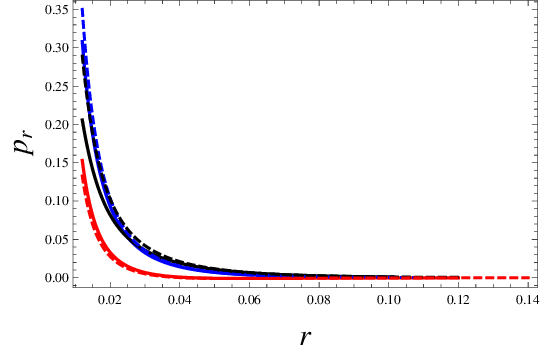,width=0.4\linewidth}
\epsfig{file=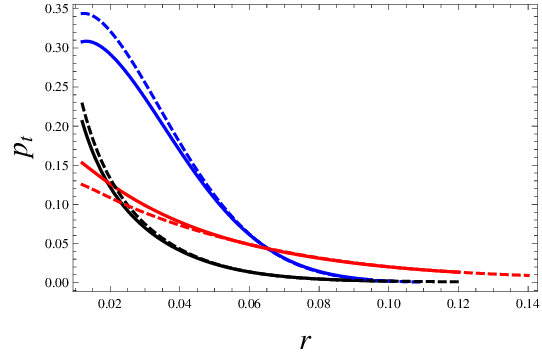,width=0.4\linewidth}\epsfig{file=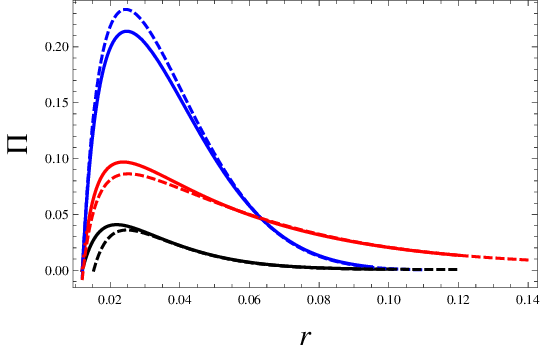,width=0.4\linewidth}
\caption{Fluid variables for $\zeta=0$
(\textcolor{blue}{\textbf{\small $\bigstar$}}), $0.1$
(\textcolor{black}{\textbf{\small $\bigstar$}}) and $0.2$
(\textcolor{red}{\textbf{\small $\bigstar$}}) for model 2 [$\chi=0$
(solid) and $0.2$ (dashed)].}
\end{figure}
\begin{figure}\center
\epsfig{file=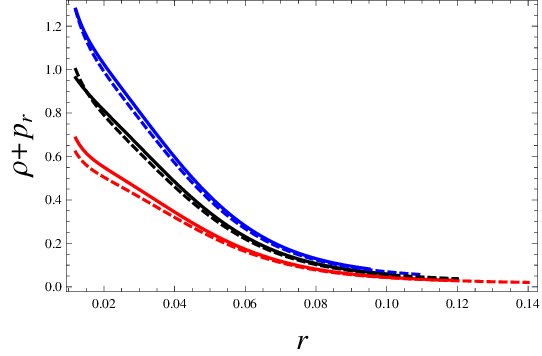,width=0.4\linewidth}\epsfig{file=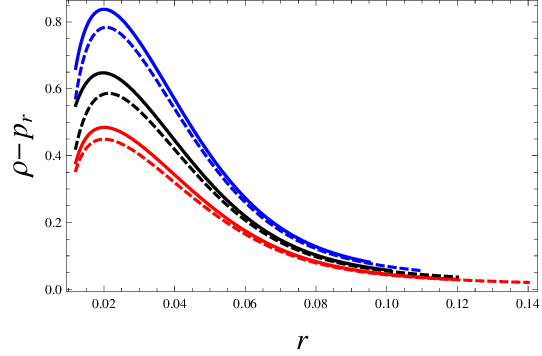,width=0.4\linewidth}
\epsfig{file=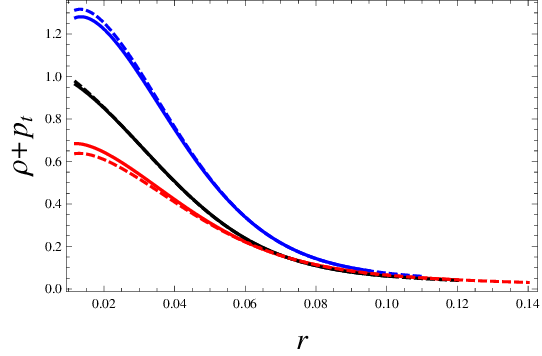,width=0.4\linewidth}\epsfig{file=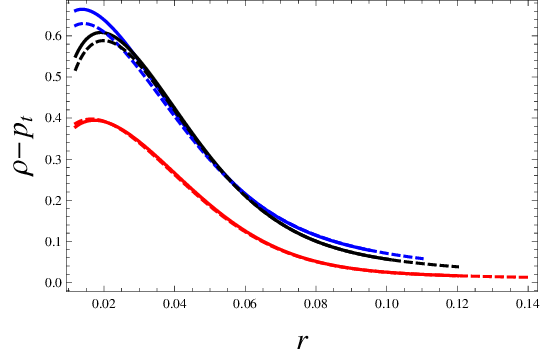,width=0.4\linewidth}
\epsfig{file=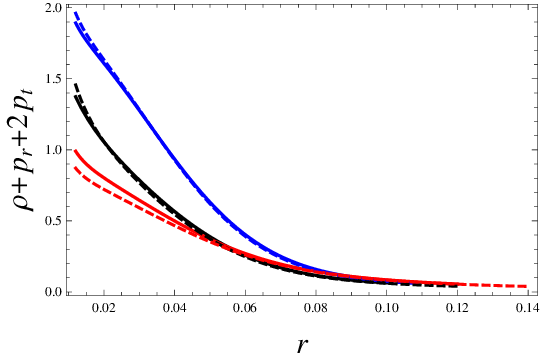,width=0.4\linewidth} \caption{Viability
for $\zeta=0$ (\textcolor{blue}{\textbf{\small $\bigstar$}}), $0.1$
(\textcolor{black}{\textbf{\small $\bigstar$}}) and $0.2$
(\textcolor{red}{\textbf{\small $\bigstar$}}) for model 2 [$\chi=0$
(solid) and $0.2$ (dashed)].}
\end{figure}
\begin{figure}\center
\epsfig{file=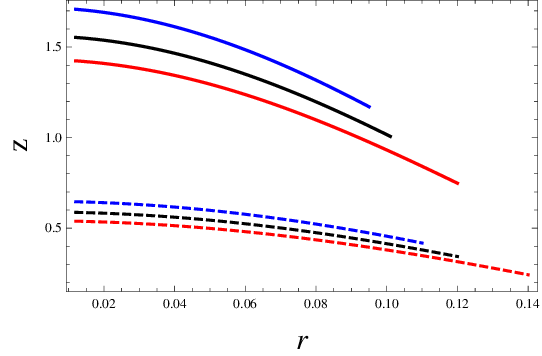,width=0.4\linewidth}\epsfig{file=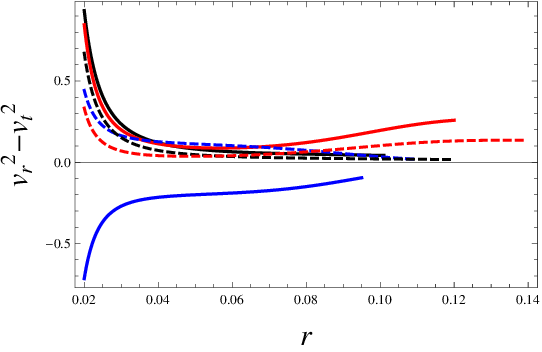,width=0.4\linewidth}
\caption{Redshift and stability for $\zeta=0$
(\textcolor{blue}{\textbf{\small $\bigstar$}}), $0.1$
(\textcolor{black}{\textbf{\small $\bigstar$}}) and $0.2$
(\textcolor{red}{\textbf{\small $\bigstar$}}) for model 2 [$\chi=0$
(solid) and $0.2$ (dashed)].}
\end{figure}

Making the differential equations to be dimensionless offers a more
efficient method for obtaining solutions. To achieve this, we
introduce novel dimensionless variables given by
\begin{align}\label{g56}
\rho_c=\frac{p_{rc}}{\tau}, \quad r=\frac{\phi}{\mathcal{B}}, \quad
\mathcal{B}^2=\frac{4\pi\rho_c}{\tau(\mathcal{U}+1)},\\\label{g57}
\Gamma^{\mathcal{X}}=\frac{\rho}{\rho_c}, \quad
\mu(\phi)=\frac{\mathcal{B}^3m(r)}{4\pi\rho_c}.
\end{align}
The central density, denoted as $\rho_{c}$, and the central radial
pressure $p_{rc}$ are pivotal components. When $r$ reaches its
maximum value at $\texttt{R}$, $\Gamma\big(\phi(\texttt{R})\big)$
becomes zero. Combining these elements with the mass \eqref{g12b}
and incorporating insights from the generalized equation detailing
evolution \eqref{g12d}, we arrive at a definitive expression as
\begin{align}\label{g58}
&\frac{d\mu}{d\phi}=2\Gamma^\mathcal{U}\phi^2\left(\frac{3\zeta-1}{4\zeta-1}\right)
-\frac{2\zeta\phi^2\Pi}{\rho_c(4\zeta-1)}
+\frac{3\zeta\tau\phi^2\Gamma^{1+\mathcal{U}}}{4\zeta-1}+\frac{2q\mathcal{B}^4}{\rho_c\phi}\frac{dq}{d\phi},\\\nonumber
&\tau(\mathcal{U}+1)\frac{d\Gamma}{d\phi}+\frac{2\Pi\Gamma^{-\mathcal{U}}}{\phi\rho_c}+\frac{\phi^3\tau\rho_c(1+\tau\Gamma)(1+\mathcal{U})}
{\mathcal{B}^2(2\tau\phi^2(1+\mathcal{U})-4\tau^2\phi(1+\mathcal{U})^2\mu+\rho_cq^2)}
\\\nonumber &\times\bigg[\tau\Gamma^{\mathcal{U}+1}\left(\frac{\zeta-1}{4\zeta-1}\right)+\frac{a\Gamma^\mathcal{U}}{4\zeta-1}
+\frac{2\Pi}{\rho_c(4\zeta-1)}+\frac{\mu}{\phi^3}-\frac{2q^2\mathcal{B}^4}{\rho_c\phi^4}\bigg]\\\label{g59}
&+\frac{1}{4\zeta-1}\bigg[\zeta
\mathcal{U}\Gamma^{-1}\frac{d\Gamma}{d\phi}-3\zeta\tau(\mathcal{U}+1)\frac{d\Gamma}{d\phi}
+\frac{2\zeta\Gamma^{-\mathcal{U}}}{\rho_c}\frac{d\Pi}{d\phi}\bigg]
-\frac{q\mathcal{B}^4\Gamma^{-\mathcal{U}}}{\phi^4\rho_c}\frac{dq}{d\phi}=0.
\end{align}

Equations \eqref{g58} and \eqref{g59} incorporate the variables
$\Gamma,~\mu$, and $\Pi$. To ensure their unique identification, an
additional constraint is required. This requirement is fulfilled by
setting $\mathbb{Y}_{TF}$ to zero. Upon expressing this condition in
terms of the variables defined in \eqref{g56} and \eqref{g57}, it
can be represented as
\begin{align}\nonumber
&16\Pi+4\phi\frac{d\Pi}{d\phi}=\frac{2\mathcal{B}^3}{\phi^3}\bigg\{10q+\frac{2\phi}{\mathcal{B}}
\bigg(2\zeta-3+\frac{\Gamma^\mathcal{U}\rho_c\phi^2}{\mathcal{B}^2}\bigg)\bigg\}
+\frac{2\rho_c\mathcal{U}\Gamma^{\mathcal{U}-1}\phi}{\mathcal{B}}+4\rho_c\Gamma^\mathcal{U}\\\label{g60}
&+\frac{12e^{-\lambda_2}\mathcal{B}^2}{\phi^2}-\frac{6e^{-\lambda_2}\mathcal{B}^2}{\phi}\frac{d\lambda_2}{d\phi}
+\zeta
e^{-\lambda_2}\bigg(\frac{3\mathcal{B}\varpi}{\phi}-\mathcal{B}\varpi\frac{d\lambda_2}{d\phi}
+\mathcal{B}\frac{d\varpi}{d\phi}-\frac{8\mathcal{B}^2}{\phi^2}+\frac{4\mathcal{B}^2}{\phi}
\frac{d\lambda_2}{d\phi}\bigg),
\end{align}
where
$$\varpi=\mathcal{B}\left(\phi\frac{d\lambda_1}{d\phi}+4\right)\left(\frac{d\lambda_2}{d\phi}
-\frac{d\lambda_1}{d\phi}\right)-2\phi\mathcal{B}\frac{d^2\lambda_1}{d\phi^2}.$$

With metric components being predetermined, the equations and
unknowns have the same number. Consequently, a solution through
numerical method is obtained admitting the conditions given below
\begin{align}\nonumber
\Gamma(\phi)|_{\phi=0}=1, \quad \Pi(\phi)|_{\phi=0}=0, \quad
\mu(\phi)|_{\phi=0}=0.
\end{align}
For $\tau = 0.1$, we analyze the trends of these variables as a
function of $\phi$ while varying the parameter $\mathcal{U}$. The
energy density, depicted in the first plot of Figure \textbf{9},
reaches its maximum at $\phi = 0$ and decreases radially, reflecting
the stable characteristics of the solution. As shown in the right
plot, the mass exhibits a consistent increase, demonstrating an
inverse correlation with $\mathcal{U}$.
\begin{figure}\center
\epsfig{file=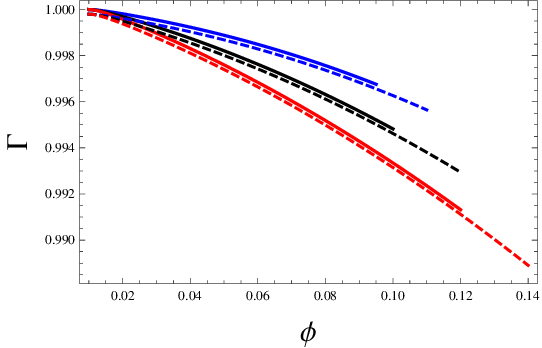,width=0.4\linewidth}\epsfig{file=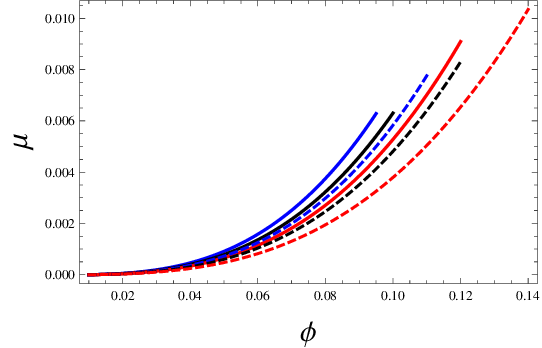,width=0.4\linewidth}
\epsfig{file=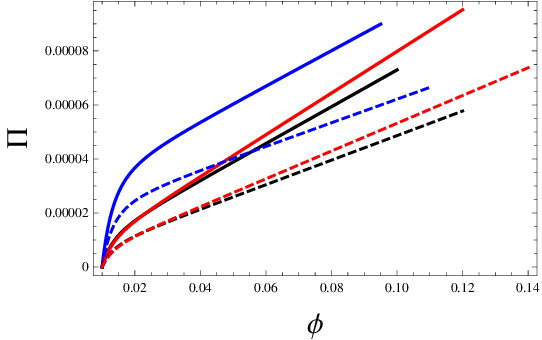,width=0.4\linewidth}
\caption{Behavior of $\Gamma,~\mu$ and $\Pi$ for $\mathcal{U}=0.06$
(\textcolor{blue}{\textbf{\small $\bigstar$}}), $0.07$
(\textcolor{black}{\textbf{\small $\bigstar$}}) and $0.08$
(\textcolor{red}{\textbf{\small $\bigstar$}}) for model 2 [$\chi=0$
(solid) and $0.2$ (dashed)].}
\end{figure}

\subsection{$\mathbb{Y}_{TF}=0$ and Non-local Equation of State}

The work by Hern\'{a}ndez and N\'{u}\~{n}ez \cite{ai} presents a
constraint connecting radial pressure to energy density through an
integral component. This term contributes to the (non-local)
equation of state, which, when paired with conditions eliminating
complexity \eqref{g30}, offers insights into intricate physical
systems. They are given as
\begin{align}\label{g62}
p_r=\rho-\frac{2}{r^3}\int_0^r\rho\bar{r}^2d\bar{r}+\frac{c_3}{2\pi
r^3}, \quad \mathbb{Y}_{TF}=0.
\end{align}
In this context, $c_3$ represents a constant term whose value could
be positive or negative. To avoid the potential singularity at the
core of the star, we set this constant to zero. The equation
presented earlier can be reinterpreted by linking it with the
interior mass \eqref{g12b}, thus providing an alternative
formulation
\begin{align}\label{g63}
&4 \chi r^6 e^{\lambda_2 }+r^2 \big[r \big\{\big(4 \zeta +\zeta r
\lambda_1 '-1\big) \big(\lambda_2 '-\lambda_1 '\big)-2 \zeta r
\lambda_1 ''\big\}+4   \zeta  \big(e^{\lambda_2 }-1\big)\big]=0.
\end{align}
The condition $\mathbb{Y}_{TF}=0$ remains applicable as outlined in
Eq.\eqref{g55b}. Consequently, both equations provide a suitable
framework for solving Eqs.\eqref{g8a}-\eqref{g10a}. Notably,
Eqs.\eqref{g55b} and \eqref{g63} incorporate higher-order metric
components, which typically require numerical methods to be solved.
This approach aligns with previous models where similar numerical
technique was employed.

Observing Figure \textbf{10}, it is evident that $e^{\lambda_1}$ and
$e^{-\lambda_2}$ exhibit non-singular and positively finite
characteristics, indicating their physical validity. Additionally,
we determine that at the center, $e^{\lambda_1(0)}$ equals a
positive constant $c_4$, while $e^{-\lambda_2(0)}$ equals unity.
Both functions converge towards a common point referred to as the
radius. A summary of these findings is presented below.
\begin{enumerate}
\item The values of radius along with the compactness factor are extracted
for $\chi=0$ as
\begin{itemize}
\item For $\zeta = 0$, the value of $\texttt{R}$ is 0.11, and the compactness
is
$$e^{\lambda_1(0.11)}=e^{-\lambda_2(0.11)}\approx0.25=1-\frac{2\texttt{M}}{\texttt{R}}
+\frac{\texttt{Q}^2}{\texttt{R}^2} \quad \Rightarrow \quad
\frac{2\texttt{M}}{\texttt{R}}-\frac{\texttt{Q}^2}{\texttt{R}^2}
\approx0.75 < 0.889.$$
\item For $\zeta = 0.1$, the value of $\texttt{R}$ is 0.12, and the compactness
is
$$e^{\lambda_1(0.12)}=e^{-\lambda_2(0.12)}\approx0.39=1-\frac{2\texttt{M}}{\texttt{R}}
+\frac{\texttt{Q}^2}{\texttt{R}^2} \quad \Rightarrow \quad
\frac{2\texttt{M}}{\texttt{R}}-\frac{\texttt{Q}^2}{\texttt{R}^2}
\approx0.61 < 0.772.$$
\item For $\zeta = 0.2$, the value of $\texttt{R}$ is 0.16, and the compactness
is
$$e^{\lambda_1(0.16)}=e^{-\lambda_2(0.16)}\approx0.48=1-\frac{2\texttt{M}}{\texttt{R}}
+\frac{\texttt{Q}^2}{\texttt{R}^2} \quad \Rightarrow \quad
\frac{2\texttt{M}}{\texttt{R}}-\frac{\texttt{Q}^2}{\texttt{R}^2}
\approx0.52 < 0.714.$$
\end{itemize}

\item On the other hand, these values for $\chi=0.2$ are
\begin{itemize}
\item For $\zeta = 0$, the value of $\texttt{R}$ is 0.13, and the compactness
is
$$e^{\lambda_1(0.13)}=e^{-\lambda_2(0.13)}\approx0.16=1-\frac{2\texttt{M}}{\texttt{R}}
+\frac{\texttt{Q}^2}{\texttt{R}^2} \quad \Rightarrow \quad
\frac{2\texttt{M}}{\texttt{R}}-\frac{\texttt{Q}^2}{\texttt{R}^2}
\approx0.84 < 0.889.$$
\item For $\zeta = 0.1$, the value of $\texttt{R}$ is 0.14, and the compactness
is
$$e^{\lambda_1(0.14)}=e^{-\lambda_2(0.14)}\approx0.29=1-\frac{2\texttt{M}}{\texttt{R}}
+\frac{\texttt{Q}^2}{\texttt{R}^2} \quad \Rightarrow \quad
\frac{2\texttt{M}}{\texttt{R}}-\frac{\texttt{Q}^2}{\texttt{R}^2}
\approx0.71 < 0.771.$$
\item For $\zeta = 0.2$, the value of $\texttt{R}$ is 0.19, and the compactness
is
$$e^{\lambda_1(0.19)}=e^{-\lambda_2(0.19)}\approx0.36=1-\frac{2\texttt{M}}{\texttt{R}}
+\frac{\texttt{Q}^2}{\texttt{R}^2} \quad \Rightarrow \quad
\frac{2\texttt{M}}{\texttt{R}}-\frac{\texttt{Q}^2}{\texttt{R}^2}
\approx0.64 < 0.711.$$
\end{itemize}
\end{enumerate}
\begin{figure}\center
\epsfig{file=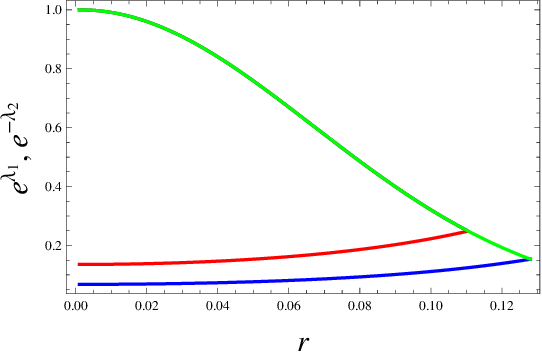,width=0.4\linewidth}\epsfig{file=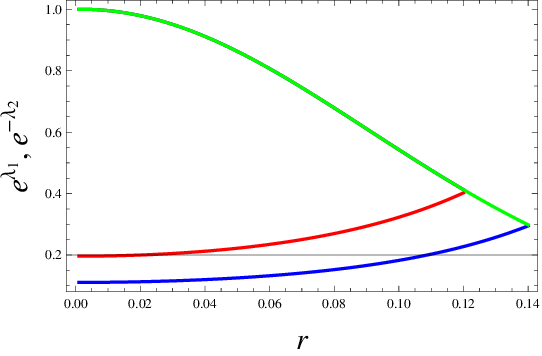,width=0.4\linewidth}
\epsfig{file=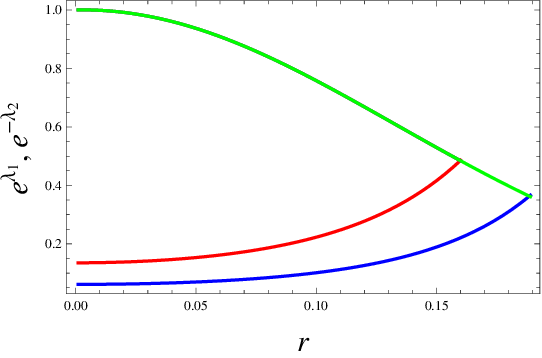,width=0.4\linewidth}
\caption{Geometric potentials $e^{\lambda_1}$
(\textcolor{red}{\textbf{\small $\bigstar$}}) ($\chi=0$) and
(\textcolor{blue}{\textbf{\small $\bigstar$}}) ($\chi=0.2$), and
$e^{-\lambda_2}$ (\textcolor{green}{\textbf{\small $\bigstar$}}) for
model 3 [$\zeta=0$ (left), $0.1$ (right) and $0.2$ (lower)]}
\end{figure}

Figure \textbf{11} illustrates the fluid triplet's compliance with
expected behavior, featuring a positive peak at its center that
gradually diminishes towards the interface for all choices of
$\zeta$ and $\chi$. Notably, the disparity between the two pressures
yields a positive anisotropy, contrasting with results from $f(R)$
theory \cite{20}. Figure \textbf{12} demonstrates adherence to
energy constraints, thereby affirming the model's viability and the
presence of conventional matter. As illustrated in Figure
\textbf{13}, the redshift exhibits a downward trend with increasing
$r$, yielding values of $1.197$, $1.008$, and $0.613$ for $\zeta$
set to $0$, $0.1$, and $0.2$, and $\chi=0$, respectively. Similarly,
these values are $0.431$, $0.361$, and $0.158$ for $\chi=0.2$. The
other plot shows that the cracking does not occur throughout the
interior domain, except when $\zeta = 0.2$. Consequently, this
solution aligns with constraints analogous to Eqs.\eqref{g34} and
\eqref{g62}, demonstrating stability for $\zeta=0$ and $\zeta=0.1$
with both values of $\chi$. This finding is consistent with previous
investigation reported in \cite{40}.
\begin{figure}\center
\epsfig{file=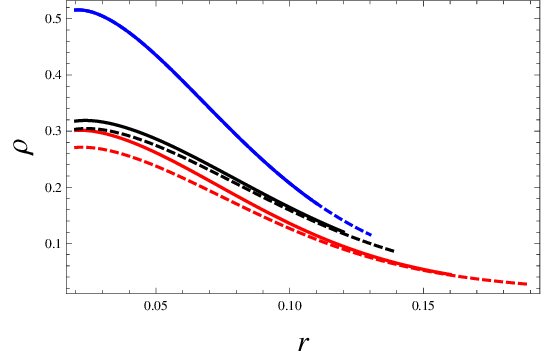,width=0.4\linewidth}\epsfig{file=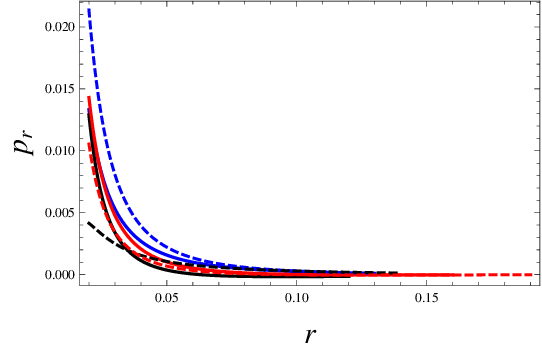,width=0.4\linewidth}
\epsfig{file=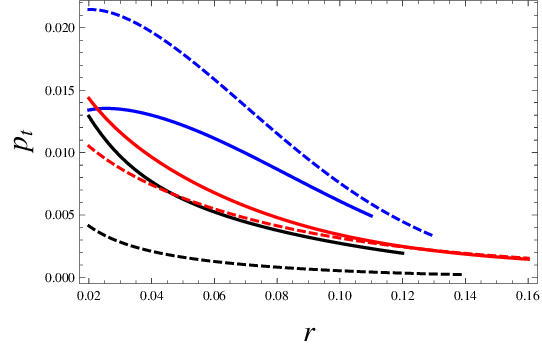,width=0.4\linewidth}\epsfig{file=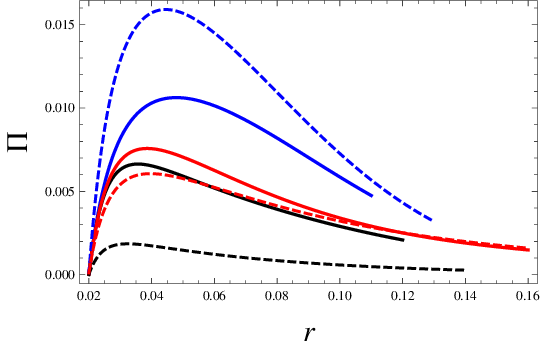,width=0.4\linewidth}
\caption{Fluid variables for $\zeta=0$
(\textcolor{blue}{\textbf{\small $\bigstar$}}), $0.1$
(\textcolor{black}{\textbf{\small $\bigstar$}}) and $0.2$
(\textcolor{red}{\textbf{\small $\bigstar$}}) for model 3 [$\chi=0$
(solid) and $0.2$ (dashed)]}
\end{figure}
\begin{figure}\center
\epsfig{file=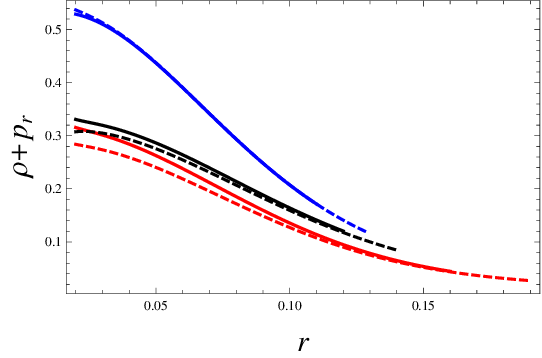,width=0.4\linewidth}\epsfig{file=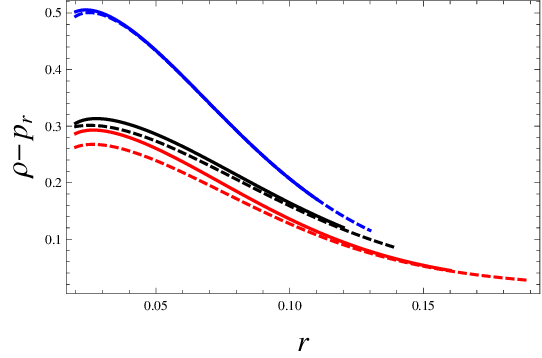,width=0.4\linewidth}
\epsfig{file=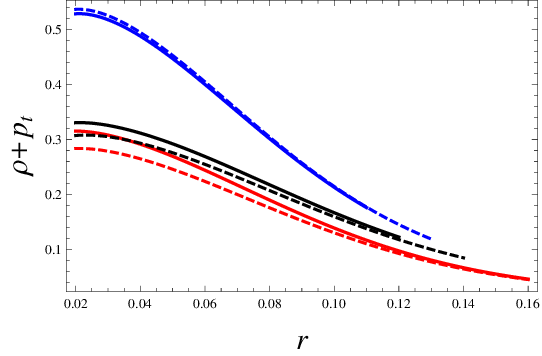,width=0.4\linewidth}\epsfig{file=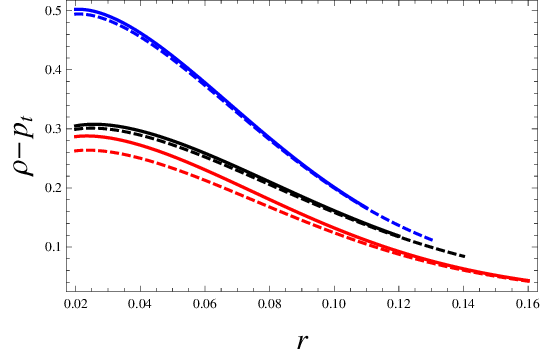,width=0.4\linewidth}
\epsfig{file=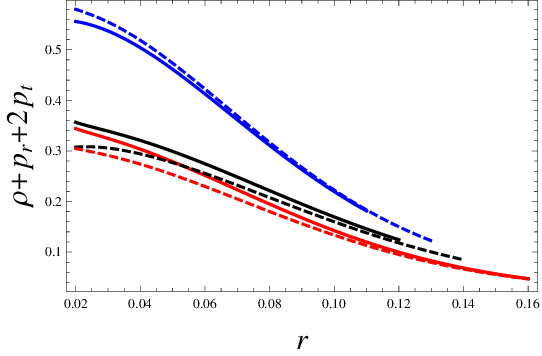,width=0.4\linewidth} \caption{Viability
for $\zeta=0$ (\textcolor{blue}{\textbf{\small $\bigstar$}}), $0.1$
(\textcolor{black}{\textbf{\small $\bigstar$}}) and $0.2$
(\textcolor{red}{\textbf{\small $\bigstar$}}) for model 3 [$\chi=0$
(solid) and $0.2$ (dashed)].}
\end{figure}
\begin{figure}\center
\epsfig{file=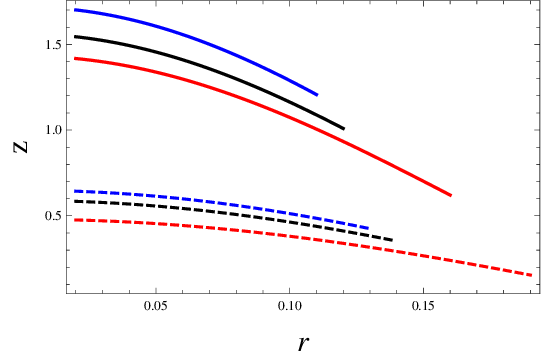,width=0.4\linewidth}\epsfig{file=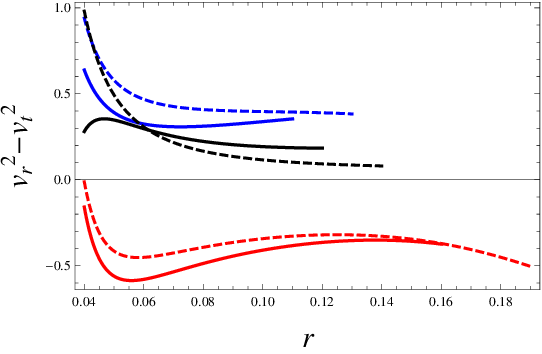,width=0.4\linewidth}
\caption{Redshift and stability for $\zeta=0$
(\textcolor{blue}{\textbf{\small $\bigstar$}}), $0.1$
(\textcolor{black}{\textbf{\small $\bigstar$}}) and $0.2$
(\textcolor{red}{\textbf{\small $\bigstar$}}) for model 3 [$\chi=0$
(solid) and $0.2$ (dashed)].}
\end{figure}

\section{Conclusions}

In this study, we have explored various solutions to the field
equations within the framework of non-conserved Rastall theory under
the influence of an electric charge. To achieve this, we have
focused on a static spherical configuration and derived the
fundamental governing equations along with a generalized evolution
equation. Furthermore, the mass function was formulated in the
context of both matter and geometric parameters. By adopting the
Schwarzschild metric as the exterior spacetime geometry, we have
been able to determine the compactness factor of such
configurations. To further proceed the analysis, we have decomposed
the curvature tensor orthogonally, yielding four distinct scalars
linked to diverse physical parameters that influence the internal
dynamics of a compact object. Notably, one scalar, denoted as
$\mathbb{Y}_{TF}$, exhibited characteristics of density
inhomogeneity and anisotropy along with Rastall corrections. This
makes it an ideal candidate for representing complexity in fluid
distribution within such objects as per the definition presented in
\cite{25}.

The set of equations \eqref{g8a}-\eqref{g10a} encompasses six
distinct unknowns: the fluid triplet, metric functions and charge.
To facilitate solving this complex system, we have implemented
specific constraints. We have chosen the charge to be in its known
form that ultimately reduces the degrees of freedom. Further, we
needed two more constraints. In this regard, the condition devoid of
complexity, as outlined in Eq.\eqref{g34}, functions as a first
constraint. Additionally, three supplementary conditions have been
applied: setting $p_r=0$, adopting a polytropic equation and a
non-local equation of state which collectively form the set of two
required constraints and result in multiple distinct models. The
challenge posed by higher-order terms in the geometric sector,
specifically $\lambda_1$ and $\lambda_2$, has been addressed by
numerically integrating the differential equations under various
feasible initial conditions. Moving forward, it is crucial to
elaborate on several key physical constraints - such as compactness,
gravitational redshift, and analysis of stability - that are vital
for developing models of compact objects.

The matter sector across the three solutions, demonstrated a
consistent pattern. This behavior peaks at the central point $(r=0)$
and gradually diminishes as the radial distance increases. Notably,
both compactness and gravitational redshift fall within acceptable
limits for all configurations. The viability criterion is satisfied
across all solutions for various parametric values as
$\zeta=0,~0.1,~0.2$ and $\chi=0,~0.2$, as evidenced by the
fulfillment of energy conditions depicted in Figures
\textbf{3},~\textbf{7}, and \textbf{12}. It must be mentioned that
the first model is not viable only for $\zeta=0$ along with
$\chi=0.2$. Furthermore, investigating potential cracking phenomena
within the interior of these models presents an intriguing avenue
for further exploration. The main results obtained from this study
are presented in the following.
\begin{itemize}
\item The condition on cracking of fluid is exclusively met by the
model corresponding to $p_r=0$, irrespective of the chosen Rastall
and charge parametric values except for $\zeta=0$ and $\chi=0.2$
(Figure \textbf{4}).

\item Stability for the second model is confined to specific
values of $\zeta$, namely $0.1$ and $0.2$ for both choices of
$\chi$. Conversely, when $\zeta=0=\chi$, instability arises,
underscoring the advantages of Rastall theory over GR (Figure
\textbf{8}). Additionally, this model also shows stable behavior
under $\zeta=0$ when charge is present.

\item For the last stellar model, stability is observed only at
$\zeta=0$ and $0.1$ along with both choices of $\chi$. On the other
hand, the cracking phenomenon appear for the rest value of $\zeta$
(Figure \textbf{13}).
\end{itemize}

It is essential to highlight that all the solutions formulated in
this study align with the findings in GR \cite{40} as well as
uncharged version of these models in Rastall theory \cite{sl5}.
Notably, the solutions corresponding to models 2 and 3 deviate from
the charged scenario already reported in \cite{30}. The results of
this investigation shall be reduced to GR for $\zeta=0=\chi$ under
uncharged scenario.
\\\\
\textbf{Data Availability:} No additional data were analyzed or
created as part of this study.

\end{document}